\documentclass[12pt,recno]{JHEP3}
\pdfoutput=1
\usepackage{graphicx,amsfonts,amssymb,stmaryrd,amsmath,pdef,longtable}
\usepackage[all]{xy}

\title{Attractor Flows from Defect Lines}  
\author{Ilka Brunner$^{1,2}$\thanks{\tt E-mail: Ilka.Brunner@physik.uni-muenchen.de} and
Daniel Roggenkamp$^{3}$\thanks{\tt E-mail: roggenka@physics.rutgers.edu} \\  
\\
${}^{1}${\small Arnold Sommerfeld Center, 
Ludwig Maximilians Universit\"at} \vspace*{-0.1cm} \\
{\small Theresienstr. 37, 
80333 M\"unchen, Germany} \vspace{0.3cm} \\
${}^{2}${\small Excellence Cluster Universe, 
Technische Universit\"at M\"unchen} \vspace*{-0.1cm} \\
{\small Boltzmannstr. 2, 
85748 Garching, Germany} \vspace{0.3cm} \\
\\
$^{3}$Department of Physics and Astronomy, Rutgers University\\
Piscataway, NJ 08855-0849, USA\\
}
\abstract{
Deforming a two dimensional conformal field theory on one side of a trivial defect line
gives rise to a defect separating the original theory from its deformation. 
The Casimir force between these defects and other defect lines or boundaries is used to construct
flows on bulk moduli spaces of CFTs. It turns out, that these flows are constant reparametrizations
of gradient flows of the $g$-functions of the chosen defect or boundary condition. The special flows associated
to supersymmetric boundary conditions in $N=(2,2)$ superconformal field theories agree with the attractor flows
studied in the context of black holes in $N=2$ supergravity. 
 }
\preprint{RUNHETC-2010-04, LMU-ASC 07/10}
\begin{document}
\section{Introduction}
Conformal defects are lines of inhomogeneity between two possibly different two dimensional conformal field theories which preserve conformal invariance. 
In this article, defect lines 
 are used to associate flows on moduli spaces of conformal field theories to given boundary conditions or defects in these theories. 
 They arise in the following way. As has been discussed in the context of Landau-Ginzburg models in \cite{Brunner:2007ur,Brunner:2008fa}, to each bulk perturbation of a conformal field theory one can associate a unique defect line between the IR and the UV theories of the corresponding RG flow. It is obtained by performing the perturbation on one side of the trivial identity defect of the UV theory. Applied to
exactly marginal perturbations, this construction yields families of defects between a given CFT and all
its deformations. 

One dimensional objects in two dimensional CFTs like defect lines or boundary conditions can however attract or repel each other. More precisely, there exists a Casimir energy and with it a force between pairs of those objects (see \eg \cite{Bachas:2001vj}). 
This applies in particular to the defects associated to deformations. Hence, extending deformations towards defect lines or boundaries can cost or yield 
energy. The amount of it depends on the defect or boundary conditions under consideration, but also on the deformation. 
Thus, any defect line or boundary condition in a given CFT defines a direction in (or rather a tangent vector to) the deformation space
of this CFT, by requiring that the  energy gain is biggest in this direction. 
If the defect or boundary condition behaves smoothly under all deformations, such that it can be carried to any point in the moduli space, then these vectors
form a vector field and hence give rise to a flow on this moduli space. To put it another way, 
in this case the Casimir energy defines a potential on the moduli space which gives rise to a gradient flow. 

Indeed, the gradient vector field can be calculated by means of perturbation theory and can be expressed purely
in terms of the coupling of the deforming fields to the chosen defect or boundary condition. Moreover,  
it turns out, that it is proportional to the gradient vector field of the logarithm of the entropy $g$-function of the
defect or boundary condition. That means that the corresponding flow is in fact a constant reparametrization of
the gradient flow of $\ln(g)$, and hence drives the bulk moduli of the CFT to local minima or saddle points of the entropy $g$. 

An interesting type of fixed points exists for flows associated to defects. 
Namely, there 
are special so called `topological' defect lines\footnote{first discussed in the context of RCFT in \cite{Petkova:2000ip}, see \cite{Frohlich:2006ch} for a recent discussion}, which have the property that correlation functions do not change when the position of these defects are changed as long as no defects or field insertions are crossed. This of course implies that there is no force between them and any other defect or boundary condition. Hence, points in the bulk moduli space, where a defect becomes topological are 
fixed points of the flow associated to this defect. Moreover, it can be shown that they are in fact attractive
fixed points, \ie minima of the $g$-function.

A related observation was already made in \cite{Bachas:2007td} in the context of the free boson CFT.
There it was noted that symmetry preserving defects in this model attract each other if and only if fusion decreases their $g$-value. Thus, these defects tend to fuse to ones with lower $g$. In that sense, defects minimizing $g$, which for the free boson are exactly
the topological defects are the most stable.

In fact, the flows derived here purely in terms of conformal field theory (at least the ones associated to supersymmetric boundary conditions), also arise in string theory, however from seemingly very different considerations. 
In the string theory terminology, the parameters of the closed string background evolve under the flow 
as to decrease the mass of chosen D-branes. Such flows are known
as `attractor flows' in the context of  BPS black holes in $N=2$ supergravity \cite{Ferrara:1995ih,Ferrara:1996um,Ferrara:1996dd,Ferrara:1997tw,Moore:1998pn}. It was noticed that the fixed points of these flows, the `attractors' correspond to arithmetically interesting geometries and at least in 
examples to rational world sheet CFTs \cite{Moore:1998pn}. 

As alluded to above, from the world sheet point of view, at least some attractor points of flows associated to defects
are points in bulk moduli space, where the defects become topological. For instance, monodromies around Gepner points
are described by topological defects (implementing the quantum symmetry) at these points of enhanced symmetry \cite{Brunner:2008fa}. Away from the Gepner
points these defects become non-topological. Therefore, Gepner points are attractors of the flows associated to the 
monodromy defects.

Another example are $N=(2,2)$ nonlinear sigma models whose target spaces exhibit `complex multiplication' (see \eg \cite{borcea}).
Complex multiplications also have a realization as topological defects in the world sheet CFTs. 
(In specific examples this can be worked out explicitly \cite{dr}.) Deforming the complex structure
of the target spaces away from complex multiplication points, these defects become non-topological. 
Hence, complex multiplication points
are attractive fixed points of the flows on the complex structure moduli spaces, which are associated to the defects. Indeed, classes of manifolds with complex multiplication have been identified as attractors in the supergravity context \cite{Moore:1998pn}, and complex multiplication has even been proposed as a criterion
for rationality of the associated world sheet CFTs \cite{Moore:1998pn,Gukov:2002nw}.

It appears that world sheet CFTs associated to general attractor points have special properties, and 
it would be very interesting to get a better understanding of their nature. 
The conformal field theory realization of attractor flows considered in this article could shed some light on this question.

Besides, it provides a natural generalization of attractor flows to non-super\-sym\-metric situations which could be of interest in string theory as well. 

Flows similar to the ones constructed here using defects 
also appear in another string theory context, namely in the treatment of backreaction of D-branes on the closed string moduli
by means of the Fischler-Susskind mechanism \cite{Fischler:1986tb,Fischler:1986ci}. There, divergences coming from higher genus string amplitudes are compensated by shifts in the coupling constants, which contribute to the RG flow of the bulk world sheet theory. 
In \cite{Keller:2007nd} this was explicitly analyzed for the divergences coming from annulus amplitudes.  
It is very interesting that the 
resulting shifted bulk RG flow is indeed closely related to the flows considered here.

The plan of the paper is as follows. In Section \ref{sec2}, a general derivation of the flows is given in terms of Casimir energies. 
It is shown that they coincide (up to a constant factor) with the gradient flows of the $g$-functions.  
In Section \ref{freeboson} the example of the free boson compactified on a circle is analyzed in detail.
For this theory, the deformation defects can be constructed exactly.
The exact expressions for their
Casimir energies are compared with the perturbative data needed in the formulation of the flows equations, and flows associated to boundary conditions and defects are discussed.
Finally, in Section \ref{n=2} we comment on the case of
$N=(2,2)$ superconformal field theories and compare to the attractor flows arising in the context of black holes in supergravity. 
\section{Deformation defects and flows on bulk moduli spaces}\label{sec2}
\FIGURE{\includegraphics[height=8cm]{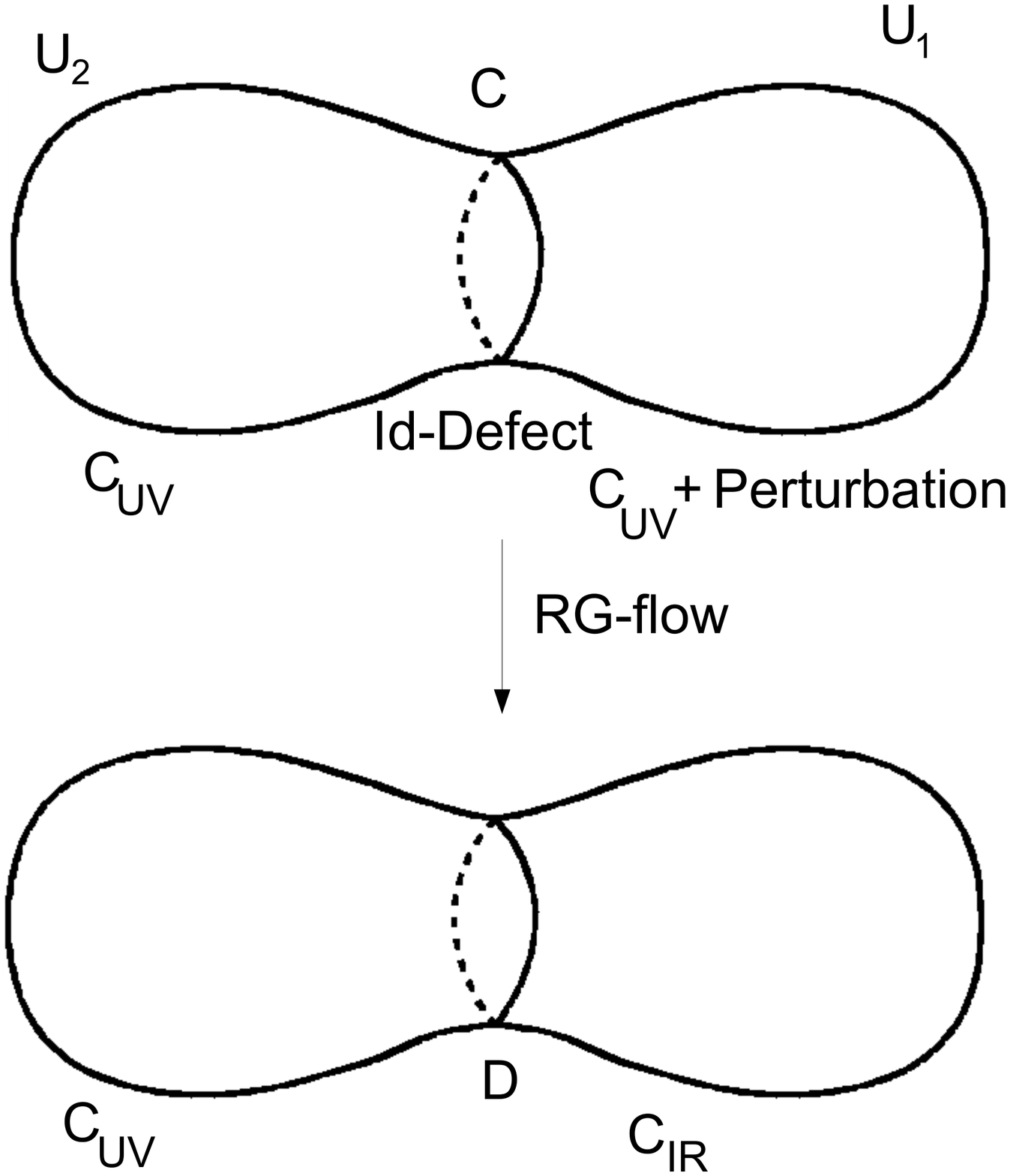}
\caption{
\small 
Perturbation on bounded domain gives rise to defect between IR- and UV-CFT.
}\label{fig1}}
As has been argued in \cite{Brunner:2007ur,Brunner:2008fa} to each bulk-pertur\-bation of a conformal
field theory, one can associate a unique 
conformal defect line between the IR and the UV fixed points of the
corresponding renormalization group flow. 
This is obtained by starting with the UV conformal field theory on a surface $\Sigma$, 
on which one places the identity defect ${\rm Id}$ (\ie the trivial defect) along a curve $C$ which cuts the surface into two domains ${\mathcal U}_1$ and ${\mathcal U}_2$. Then one perturbs the conformal field theory, but only on one of these domains ${\mathcal U}_1$
\beq
\bra\cdots\ket_{\lambda}=\bra \cdots e^{\Delta S}\ket\,,\\
\quad\Delta S=\sum_i\lambda^i\int_{{\mathcal U}_1}{{\rm d}^2z\over -2i}\varphi_i(z,\ol{z})\,.\nonumber
\eeq

The end point of the corresponding renormalization group flow is given by 
the IR CFT on the domain ${\mathcal U}_1$ separated by a non-trivial conformal defect from the UV CFT on ${\mathcal U}_2$. 
Thus,
given any bulk flow between two CFTs one obtains in this way a conformal defect line between the IR and UV CFT (\cf figure \ref{fig1}). 
Since the original identity defect is transparent in particular to the perturbing fields, no additional regularization is needed on this defect. Therefore, for a given bulk flow this defect is uniquely defined.

Such flow defects in particular exist for exactly marginal perturbations, 
giving rise to families ${\mathcal D}_\lambda$ of conformal defects between 
a given CFT and all its deformations parametrized by $\lambda$.  

Now, one dimensional objects in CFTs such as defect lines or 
boundary conditions repel or attract each other. 
More precisely, there is a Casimir energy associated to any pair 
of such objects \cite{Bachas:2001vj}.
This energy can be expressed by means of cylinder amplitudes. 
Let us first discuss the Casimir energy between a defect line ${\mathcal D}$ 
and a boundary condition $B$. 
\FIGURE{
\includegraphics[height=3.6cm]{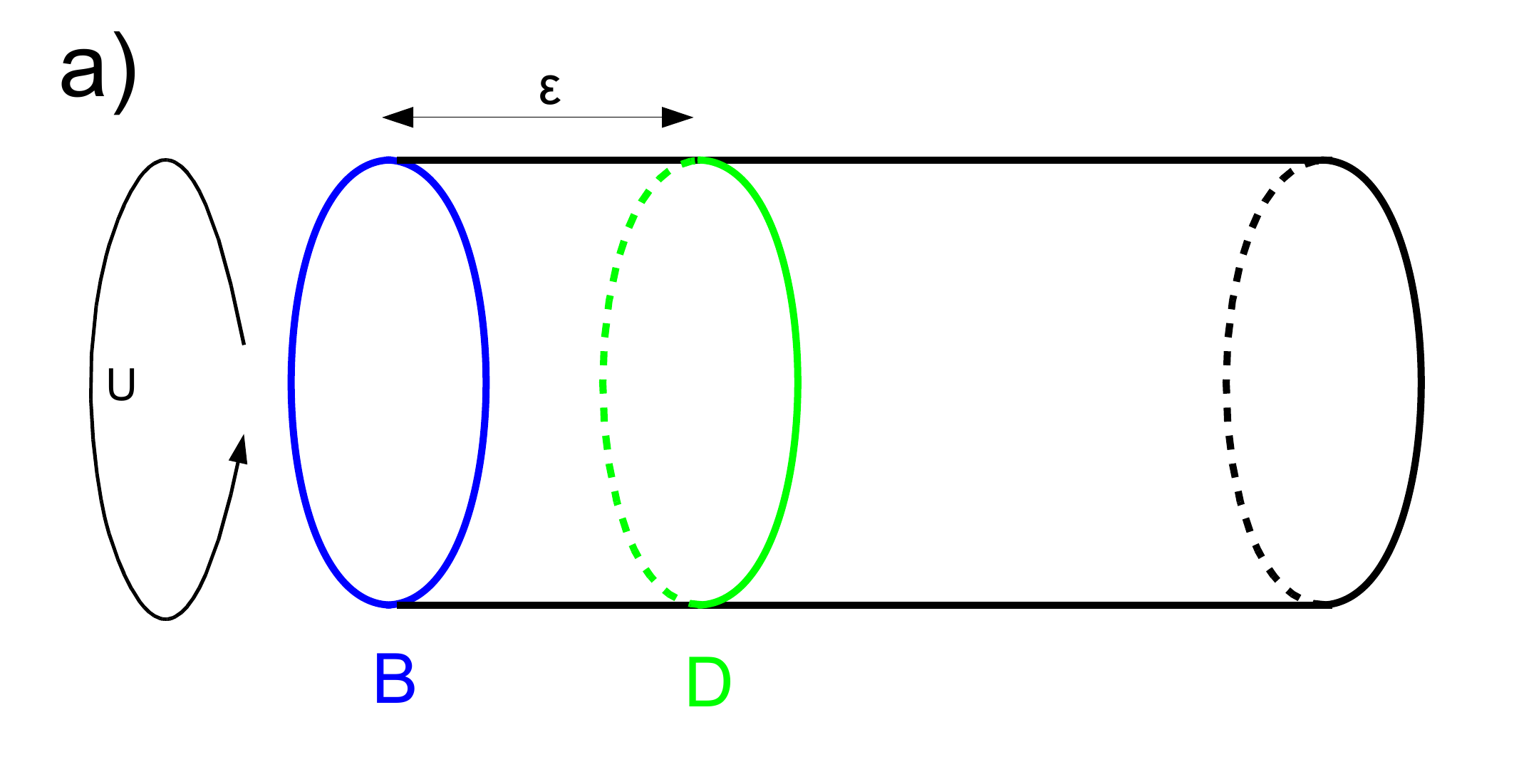}
\includegraphics[height=3.6cm]{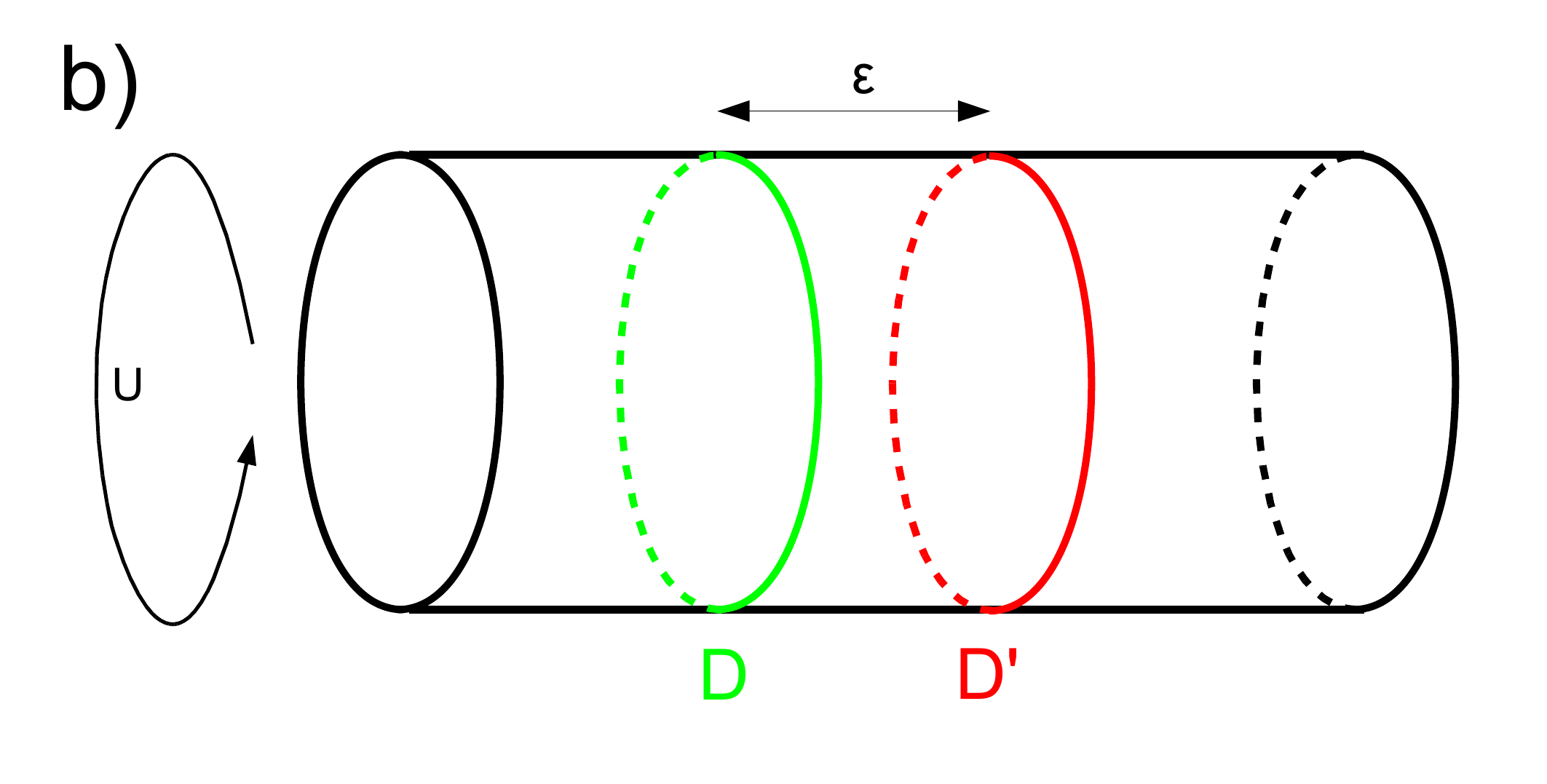}
\caption{
\small Cylinder amplitudes: a) ${\mathcal B}_{B,{\mathcal D}}^{\epsilon, U}$, and b) ${\mathcal B}_{{\mathcal D},{\mathcal D}\p}^{\epsilon, U}$.
}\label{fig2}
}
The amplitude on a half infinite cylinder 
with boundary condition $B$ imposed on the finite boundary, defect line ${\mathcal D}$ 
placed at distance $\epsilon$ parallel to the boundary, and the vacuum $\Omega$ 
inserted at the infinite end of the cylinder (\cf figure \ref{fig2}a)
can be regarded in two dual ways, namely as
vacuum transmitted through the defect and absorbed on the boundary, 
or as loop of half open defect twisted states:
\beq
{\mathcal B}_{B,{\mathcal D}}^{\epsilon,U}=\brai B\|e^{-2\pi{\epsilon\over U} H}{\mathcal D}|\Omega\ket
={\rm tr}_{\HH_{B,{\mathcal D}}}\left(e^{-2\pi{U\over\epsilon}H_{B,{\mathcal D}}}\right)\,.
\eeq
Here $U$ is the circumference of the cylinder, $2\pi H$ the bulk Hamiltonian of the CFT, $\HH_{B,{\mathcal D}}$ is the Hilbert space of half open twisted states, and $2\pi H_{B,{\mathcal D}}$ is the Hamiltonian on this space. 
The ground state energy on $\HH_{B,{\mathcal D}}$ can then be obtained as the limit
\beq
{\mathcal E}_{B,{\mathcal D}}^\epsilon={{\mathcal E}_{B,{\mathcal D}}\over\epsilon}=-\lim_{U\to\infty}{1\over U}\ln\left({\mathcal B}_{B,{\mathcal D}}^{\epsilon,U}\right)\,,
\eeq
where ${\mathcal E}_{B,{\mathcal D}}$ is the smallest eigenvalue of $2\pi H_{B,{\mathcal D}}$. 
This Casimir energy gives rise to a force
\beq
{\mathcal F}_{B,{\mathcal D}}^\epsilon=-{\partial\over\partial\epsilon}{\mathcal E}_{B,{\mathcal D}}^\epsilon={{\mathcal E}_{B,{\mathcal D}}\over\epsilon^2}
\eeq
between defect and boundary. 

If the defect ${\mathcal D}$ is topological, then it commutes with the Hamiltonian, and in particular ${\mathcal B}_{B,{\mathcal D}}$ is independent of $U$. Therefore, the Casimir energy is zero, and there is no force between defect and boundary. 
General conformal defects however do not commute with $H$ and are attracted or repelled by boundaries.
This is true in particular 
for the defects associated to deformations of CFTs. Hence, any boundary condition
in a given CFT
defines a direction in the deformation space of this CFT 
simply by the condition that the deformation defect associated to this 
direction is the deformation defect attracted the most by the boundary condition. 
This direction is given by the gradient 
\beq\label{grad}
-\nabla_{\lambda}\Big|_{\lambda=0}{\mathcal E}_{B,{\mathcal D_\lambda}}\,,
\eeq
where $\lambda^i$ are local coordinates on the moduli space.  

If the boundary condition $B$ is smoothly deformed to $B(\lambda)$ along the bulk deformations\footnote{This is the case if the bulk deformation does not trigger a relevant flow in the boundary sectors, \ie
the bulk-boundary OPE of the deforming bulk fields $\varphi_j$ does not give rise to relevant boundary
fields.}, 
then it defines in this way a vector field and with it a flow on the moduli space of the bulk CFT. 
Denoting by ${\mathcal D}_{\lambda,\widetilde{\lambda}}$ the deformation defect from theory ${\widetilde{\lambda}}$ 
to theory $\lambda$ this flow can be written as
\beq
{{\rm d}\over{\rm d}t}\lambda^i=-g^{ij}{\partial\over\partial\widetilde{\lambda}^j}\Big|_{\widetilde{\lambda}=\lambda}{\mathcal E}_{B(\lambda),{\mathcal D}_{\lambda,\widetilde{\lambda}}}\,.
\eeq
Here $g^{ij}$ is the inverse of the Zamolodchikov metric $g_{ij}=\bra \varphi_i|\varphi_j\ket$ 
on the moduli space. 

For deformation defects the gradient \eq{grad} can of course be calculated by means of perturbation
theory
\beq
\partial_{\lambda^j}\Big|_{\lambda=0}{\mathcal E}_{B,{\mathcal D}_\lambda}=
-\lim_{U\to\infty}{\epsilon\over  U}{\partial_{\lambda^j}{\mathcal B}_{B,{\mathcal D}_\lambda}^{\epsilon, U}\over {\mathcal B}_{B,{\mathcal D}_\lambda}}\Big|_{\lambda=0}\,,
\eeq
where 
\beq
\partial_{\lambda^j}\Big|_{\lambda=0}{\mathcal B}_{B,{\mathcal D}_\lambda}^{\epsilon, U}
=\int_{C_{\epsilon,U}}{{\rm d}^2z\over -2i}\brai B\|\varphi_j(z,\ol{z})|\Omega\ket\,,
\eeq
is the first order term in the perturbative expansion of the amplitude ${\mathcal B}_{B,{\rm Id}}^{\epsilon,U}$
 perturbed by the field $\varphi_j$ on the side of the identity defect which is opposite to the boundary. That means, 
$C_{\epsilon,U}$ is the half infinite cylinder of circumference $U$ excluding 
the strip of width $\epsilon$ between
the defect and the boundary.
\FIGURE{\includegraphics[height=5cm]{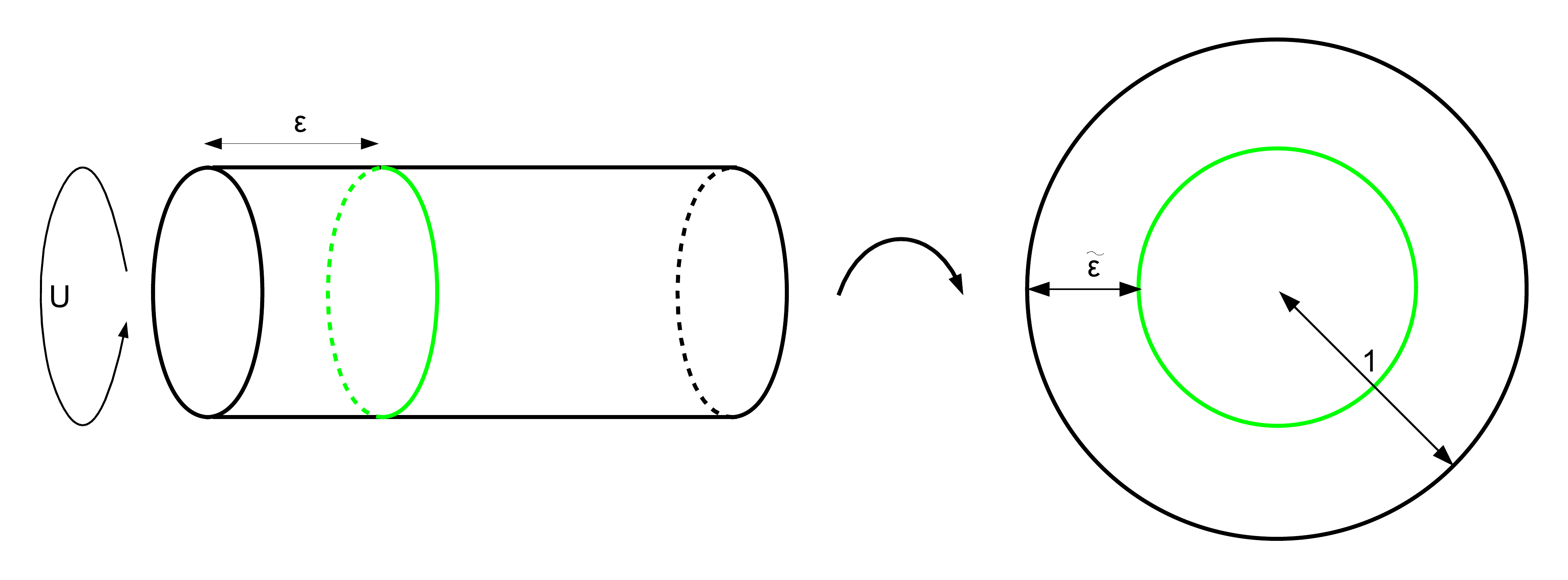}
\caption{
\small Mapping the cylinder to the unit disk.
}\label{fig3}}
This can be more conveniently calculated on the disk, so we map the half infinite cylinder to the unit disk $D=D_1$ by $z\mapsto e^{2\pi z\over U}$ (\cf figure \ref{fig3}), 
under which 
\beq
C_{\epsilon,U}\mapsto D_{1-\widetilde\epsilon}\,,\quad 1-\widetilde\epsilon=e^{-2\pi{\epsilon\over U}}\,.
\eeq
Since the $\varphi_j$ are marginal, \ie have conformal weights $h_j=\ol{h}_j=1$, the correlation function on the disk is given by
\beq\label{onepointfunction}
\brai B\|\varphi_j(z,\ol{z})|\Omega\ket={g_BB_{\varphi_j}^{(B)}\over(1-|z|^2)^2}\,,
\eeq
where $g_B=\brai B\|\Omega\ket$ is the $g$-factor of the boundary condition $B$ and 
$B_{\varphi_j}^{(B)}:=B_{\varphi_j}^{(B)\;1}$ is the bulk-boundary OPE coefficient of the perturbing bulk
field $\varphi_j$ and the identiy $1_B$ on the boundary $B$. 
The integral of this correlation function can be easily computed
\beqn\label{integral}
\int_{D_{1-\widetilde{\epsilon}}}{{\rm d}^2z\over -2i}{1\over (1-|z|^2)^2}&=&
\pi\int_0^{(1-\widetilde{\epsilon})^2}{\rm d}x (1-x)^{-2}=\pi{(1-\widetilde{\epsilon})^2\over1-(1-\widetilde{\epsilon})^2}\\
&=&\pi{e^{-4\pi{\epsilon\over U}}\over 1-e^{-4\pi{\epsilon\over U}}}\,.\nonumber
\eeqn
Since ${\mathcal D}_0={\rm Id}$ is the identity defect, and
\beq
\lim_{U\to\infty}{\epsilon\over U}\pi 
{e^{-4\pi{\epsilon\over U}}\over 1-e^{-4\pi{\epsilon\over U}}}
={1\over 4}\,,
\eeq
one finds for the gradient
\beq\label{generalgrad}
\partial_{\lambda_j}\Big|_{\lambda=0}{\mathcal E}_{B,{\mathcal D}_\lambda}=
-{B_{\varphi_j}^{(B)}\over 4}\,.
\eeq
Thus, the flow on bulk moduli space associated to the boundary condition
$B$ can be expressed as
\beq\label{generalflow}
{{\rm d}\over {\rm d}t}\lambda^i={1\over 4}g^{ij}B_{\varphi_j}^{(B)}\,.
\eeq
The fixed points of this flow are those deformations of the bulk theory, in which the perturbing
fields do not couple to the chosen boundary condition $B$. 

Interestingly, up to a factor ${1\over 2\pi}$, \ie a constant reparametrization this is nothing but the
gradient flow of the logarithm $\ln(g_{B})$ of the $g$-function of $B$. 
This can be seen by means of first order perturbation theory as follows
\beq
\partial_{\lambda^j}\Big|_{\lambda=0}\ln(g_{B})=
{1\over g_{B}}\int_{D_{1-\widetilde{\epsilon}}}{{\rm d}^2z\over -2i}\brai B\|\varphi_j(z,\ol{z})|0\ket\,,
\eeq
where again we have mapped the perturbation from the semi-infinite cylinder to the unit disk. 
Note however, that here the parameter $\epsilon$ serves as regularization parameter. 
The relevant one point function on the disk has been given in \eq{onepointfunction} above, 
and its
integral has been calculated in 
\eq{integral}. Expanding in ${\epsilon\over U}$, one obtains
\beq
\int_{D_{1-\widetilde{\epsilon}}}{{\rm d}^2z\over -2i}{1\over (1-|z|^2)^2}={U\over 4\epsilon}-{\pi\over 2}+O\left({\epsilon\over U}\right)\,.
\eeq
In the minimal subraction scheme, renormalization of fields and coupling constants exactly cancels
singular terms proportional to $\ln(\epsilon)$ and $\epsilon^{-n}$, $n>0$. Hence, the remaining finite part is 
\beq
\partial_{\lambda^j}\Big|_{\lambda=0}\ln(g_{B})=-{\pi\over 2}B_{\varphi_j}^{(B)}\,,
\eeq
and the gradient flow of $\ln(g_{B})$ is given by
\beq\label{lngflow}
{{\rm d}\over{\rm d}t}\lambda^i
=-g^{ij}\partial_{\lambda^j}\ln(g_B)
={\pi\over 2}g^{ij}B_{\varphi_j}^{(B)}\,.
\eeq
Comparing with \eq{generalgrad} we see that the flow obtained by means of the Casimir energy of deformation
 defects is a constant reparametrization of the gradient flow of $\ln(g_B)$, and hence the flow associated to a 
 boundary condition $B$ decreases its $g$-factor.

As alluded to above, defects are not only attracted or repelled by boundaries but also by other defect lines.
Hence, in the same way as boundary conditions also defect lines give rise to flows on bulk moduli spaces. The Casimir 
energy between two defect lines can be obtained analogously to the boundary case as a limit of a
cylinder correlation function ${\mathcal B}_{{\mathcal D},{\mathcal D}\p}^{\epsilon, U}$. 
In this case however, the cylinder extends to infinity in both directions with the vacuum
inserted at both ends. The two defect lines are placed parallel at distance $\epsilon$ around it (\cf figure \ref{fig2}b).
Analogously to the boundary case, 
this correlation function can be interpreted in two different ways, as the transmission of the vacuum
through the two defects or a trace over a Hilbert space $\HH_{{\mathcal D},{\mathcal D}\p}$ of defect twisted states
\beq
{\mathcal B}_{{\mathcal D},{\mathcal D}\p}^{\epsilon,U}=\bra\Omega|{\mathcal D}e^{-2\pi{\epsilon\over U} H}{\mathcal D}\p|\Omega\ket
={\rm tr}_{\HH_{{\mathcal D},{\mathcal D}\p}}\left(e^{-2\pi{U\over\epsilon}H_{{\mathcal D},{\mathcal D}\p}}\right)\,,
\eeq
giving rise to the Casimir energy 
\beq
{\mathcal E}_{{\mathcal D},{\mathcal D}\p}^\epsilon={{\mathcal E}_{{\mathcal D},{\mathcal D}\p}\over\epsilon}=-\lim_{U\to\infty}{1\over  U}\ln\left({\mathcal B}_{{\mathcal D},{\mathcal D}\p}^{\epsilon,U}\right)\,.
\eeq
With the same reasoning as for boundary conditions, the Casimir energy between any defect ${\mathcal D}$
and a deformation defect ${\mathcal D}\p={\mathcal D}_\lambda$ defines a direction in the deformation space
of the conformal field theory on one side of the defect line ${\mathcal D}$
\beq
-\nabla_{\lambda}\Big|_{\lambda=0}{\mathcal E}_{{\mathcal D},{\mathcal D_\lambda}}\,. 
\eeq
Hence, if ${\mathcal D}$ behaves smoothly under the corresponding deformations\footnote{As for boundary conditions this is the case if the bulk deformation does not trigger a relevant flow in the defect sectors, \ie the bulk-defect OPE of the deforming bulk fields $\varphi_j$ does not produce relevant defect fields.} it gives rise to a vector field and hence flow on the moduli space of 
the bulk CFT on one of its sides
\beq
{{\rm d}\over{\rm d}t}\lambda^i=-g^{ij}{\partial\over\partial\widetilde{\lambda}^j}\Big|_{\widetilde{\lambda}=\lambda}{\mathcal E}_{{\mathcal D}(\lambda),{\mathcal D}_{\lambda,\widetilde{\lambda}}}\,. 
\eeq
By means of perturbation theory, this flow can be rewritten as
\beq\label{generalgraddef}
{{\rm d}\over {\rm d}t}\lambda^i
={1\over 4}g^{ij}B_{\varphi_j}^{({\mathcal D})}\,,
\eeq
where $B_{\varphi_j}^{({\mathcal D})}$ is the OPE coefficient of the perturbing field with the 
identity on the defect ${\mathcal D}$. This is completely analogous to the boundary case. In fact, the
perturbative calculation in the defect case can be reduced to that for boundaries by means of the folding 
trick\footnote{The folding trick maps a correlation function on a cylinder with a defect inserted in the middle to the correlation function on the half cylinder, where one of the halves is folded over to the other one with a respective boundary condition imposed on the boundary thus created \cite{WOng:1994pa,Oshikawa:1996ww,Bachas:2001vj}. 
Therefore, the perturbative analysis including bulk-one-point functions is completely analogous.}.
This implies that the flow \eq{generalgraddef} is indeed also a constant reparametrization of
the gradient flow 
\beq\label{lngflowdef}
{{\rm d}\over{\rm d}t}\lambda^i
=-g^{ij}\partial_{\lambda^j}\ln(g_{\mathcal D})
={\pi\over 2}g^{ij}B_{\varphi_j}^{({\mathcal D})}\,.
\eeq
of the logarithm of the defect $g$-function defined by
\beq
g_{\mathcal D}=\bra\Omega|{\mathcal D}|\Omega\ket\,.
\eeq
Note that points $\lambda$, in which ${\mathcal D}$ is a topological defect are fixed points of this flow. 
This follows from the fact that the Casimir energy between a topological defect and any other defect is zero. 
In the perturbative approach, bulk-one-point functions in the presence of a topological defect 
vanish due to conformal covariance, so that $B_{\varphi_j}^{({\mathcal D})}=0$. 

It can be shown that points
where ${\mathcal D}$ is a topological defect
are in fact attractive fixed points, \ie local minima of $g_{\mathcal D}$. As is calculated in Appendix \ref{Hessian}, 
the Hessian of $\ln(g_{\mathcal D})$ at such a point is given by
\beq
\partial_{\lambda^i}\partial_{\lambda^j}\Big|_{\lambda=0}\ln(g_{\mathcal D})=
{\pi^2\over 4} g_{ij}\,,
\eeq
which is positive definite.
Note however that the defect ${\mathcal D}$ is not necessarily topological in all fixed points of the flow
Extreme examples of non-topological fixed points are purely reflective defects. These do not transmit any excitations and hence consist of boundary conditions for the theories on each side. In particular they are not topological. The flows
associated to such defects are given by the flows associated to the respective boundary conditions, and the fixed points are again purely reflective defects.

%
\FIGURE{\includegraphics[height=4.4cm]{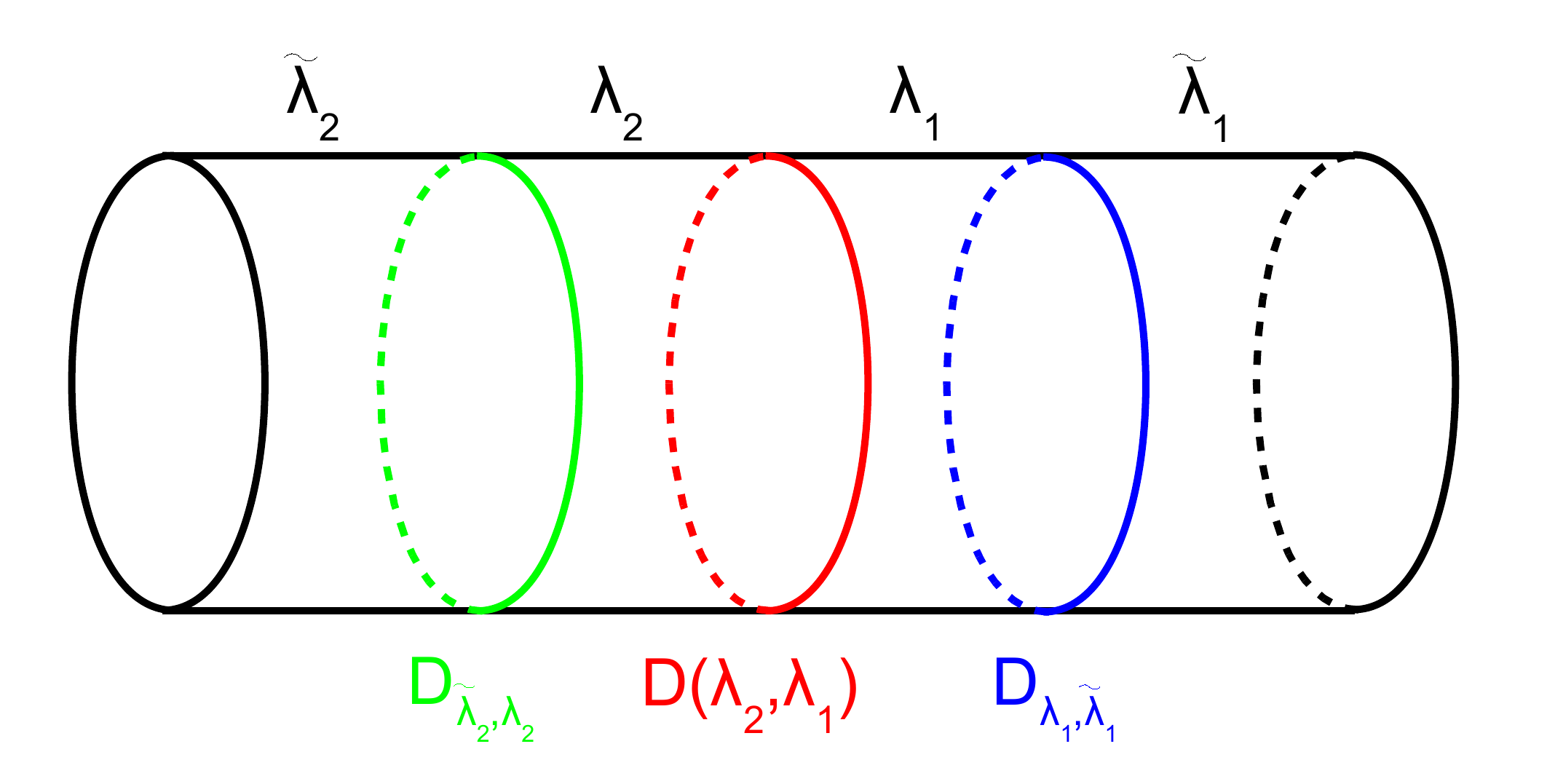}
\caption{
\small 
Cylinder with defect ${\mathcal D}$ in between two deformation defects.
}\label{fig4}}
%
Of course, one can also consider deformations on both sides of a defect line ${\mathcal D}$. The relevant 
Casimir energy can then be expressed in terms of a cylinder amplitude with three defects parallel to each other,
the two deformation defects, and ${\mathcal D}$ in between (\cf figure \ref{fig4}).  
The corresponding coupled gradient flow is given by
\beqn
{{\rm d}\over{\rm d}t}\lambda^i_1&=&-g^{ij}_1{\partial\over\partial\widetilde{\lambda}^j_1}\Big|_{\widetilde{\lambda}_i=\lambda_i}
{\mathcal E}_{{\mathcal D}_{\widetilde{\lambda}_2,\lambda_2},{\mathcal D}(\lambda_2,\lambda_1),{\mathcal D}_{\lambda_1,\widetilde{\lambda}_1}}\,,\\
 {{\rm d}\over{\rm d}t}\lambda^i_2&=&-g^{ij}_2{\partial\over\partial\widetilde{\lambda}^j_2}\Big|_{\widetilde{\lambda}_i=\lambda_i}
{\mathcal E}_{{\mathcal D}_{\widetilde{\lambda}_2,\lambda_2},{\mathcal D}(\lambda_2,\lambda_1),{\mathcal D}_{\lambda_1,\widetilde{\lambda}_1}}\,,\nonumber
\eeqn
where $\lambda_1^i$, $\widetilde{\lambda}_1^i$ are the coupling constants on one side of the defect, and
$\lambda_2^i$, $\widetilde{\lambda}_2^i$ the ones on the other.  First order deformation theory 
yields
\beq
{{\rm d}\over {\rm d}t}\lambda^i_1
={1\over 4}g^{ij}_1B_{\varphi_j^1}^{({\mathcal D})}\,,\quad
{{\rm d}\over {\rm d}t}\lambda^i_2
={1\over 4}g^{ij}_2B_{\varphi_j^2}^{({\mathcal D})}\,.
\eeq
Again, this is a constant reparametrization of the gradient flow of $\ln(g_{\mathcal D})$. 

If ${\mathcal D}$ is a defect between one and the same theory, then it makes sense to consider 
simultaneous deformations on both sides, so as to keep it a defect between one and the same theory. 
The gradient flow is then given by
\beq
{{\rm d}\over{\rm d}t}\lambda^i=-g^{ij}{\partial\over\partial\widetilde{\lambda}^j}\Big|_{\widetilde{\lambda}=\lambda}
{\mathcal E}_{{\mathcal D}_{\widetilde{\lambda},\lambda},{\mathcal D}(\lambda),{\mathcal D}_{\lambda,\widetilde{\lambda}}}\,.
\eeq
In first order one picks up perturbations on both sides of ${\mathcal D}$ and one obtains
\beq\label{gen2flow}
{{\rm d}\over {\rm d}t}\lambda^i
={1\over 4}g^{ij}\left(B_{\varphi_j^1}^{({\mathcal D})}+B_{\varphi_j^2}^{({\mathcal D})}\right)\,,
\eeq
where $\varphi_j^1$ and $\varphi_j^2$ refer to the perturbing fields on the two sides of
the defect line. By the same arguments as before this is a constant reparametrization of the gradient flow of
$\ln(g_{\mathcal D})$, where here the bulk moduli are varied on both sides of the defect simultaneously. 
\section{Example:  The free boson}\label{freeboson}

As an example we consider the free boson compactified on a circle. 
This conformal field theory is governed by a holomorphic and an antiholomorphic 
$\widehat{\mathfrak u}(1)$ current algebra which are generated by currents
\beq
j(z)=\sum_{n}a_nz^{-n-1}\,,\qquad \ol{\jmath}(\ol{z})=\sum_n\ol{a}_n\ol{z}^{-n-1}\,.
\eeq
In terms of the currents, the energy momentum tensor can be expressed as
\beq
T={1\over 2}:jj:\,,\qquad\ol{T}={1\over 2}:\ol{\jmath}\ol{\jmath}:\,.
\eeq
The Hilbert space of the theory decomposes into the respective highest weight modules 
$\VV_Q\otimes\ol{\VV}_{\ol{Q}}$ of ${\mathfrak u}(1)\oplus\ol{{\mathfrak u}(1)}$-charge $(Q,\ol{Q})$.
The corresponding highest weight vectors $|(Q,\ol{Q})\ket$ have conformal weights $(h,\ol{h})={1\over 2}(Q^2,\ol{Q}^2)$.  The model is completely determined by the lattice
$\Gamma$ of ${\mathfrak u}(1)\oplus\ol{{\mathfrak{u}}(1)}$ charges appearing in the theory
\beq
\HH_\Gamma=\bigoplus_{(Q,\ol{Q})\in\Gamma} \VV_Q\otimes \ol{\VV}_{\ol Q}\,.
\eeq
$\Gamma$ is an even selfdual lattice in $\RR^{1,1}$ and it comes in a family
\beq
\widetilde{{\mathcal M}}={\rm O}(1,1)/{\rm O}(1)\times {\rm O(1)}\cong\RR^+\,.
\eeq
The latter can be parametrized as
\beq
\Gamma_R=\left\{{1\over\sqrt{2}}\left({n\over R}+mR,{n\over R}-mR\right)\,\big|\,n,m\in\ZZ\right\}\,.
\eeq
Here $R\in\RR^+$ can be regarded as the radius of the target space circle. The lattices $\Gamma_R$ for different $R$ are mapped to each other by means of ${\rm O}(1,1)$-transformations, explicitly one has
\beq\label{Ofam}
\Gamma_{R\p}={\mathcal O}(\theta)\Gamma_R\,,\qquad
{\mathcal O}(\theta)=\left(\begin{array}{cc} \cosh(\theta)&\sinh(\theta)\\\sinh(\theta)&\cosh(\theta)\end{array}\right)\in {\rm O}(1,1)\,,\quad e^\theta={R\over R\p}\,.
\eeq
Thus, there is a one-parameter
family of free bosonic conformal field theories para\-me\-trized by the radius of the 
target space circle. Indeed, this family can be generated by exactly marginal perturbations
with the operator $\varphi=j\ol{\jmath}$ which preserve the $\widehat{\mathfrak u}(1)\oplus\ol{\widehat{\mathfrak u}}(1)$-current algebra (see \eg \cite{chaud}).
Perturbing a theory with circle radius $R_0$ by adding 
\beq
\Delta S=\lambda\int_{\Sigma}{\rm d}^2z j(z)\ol{\jmath}(\ol{z})
\eeq
to the action one obtains the circle theory with radius
\beq
R(\lambda)=R(0)e^{\pi\lambda}\,.
\eeq
This can be seen by comparing the variation of ${\mathfrak u}(1)\oplus\ol{\mathfrak u}(1)$-charges
under the perturbation with the $R$-dependence of the charges 
\beq
(Q^R_{n,m},\ol{Q}_{n,m}^R)={1\over\sqrt{2}}\left({n\over R}+mR,{n\over R}-mR\right)\,.
\eeq
Namely, by perturbation analysis one finds (see Appendix \ref{pertapp})
\beq 
\partial_{\lambda} Q(\lambda)=-\pi\ol{Q}(\lambda)\,,\quad
\partial_{\lambda} \ol{Q}(\lambda)=-\pi{Q}(\lambda)\,,
\eeq
whereas
\beq
\partial_{\ln(R)} Q_{n,m}^R=-\ol{Q}_{n,m}^R\,,\quad
\partial_{\ln(R)} \ol{Q}_{n,m}^R=-{Q}_{n,m}^R\,,
\eeq
and hence 
\beq
{\partial\ln(R)\over\partial\lambda}=\pi\,.
\eeq
\subsection*{Symmetry-preserving defects}
In this Section we describe the $\widehat{\mathfrak u}(1)\oplus\ol{\widehat{\mathfrak u}}(1)$-preserving defects between possibly different CFTs in the free boson family (\cf \cite{Bachas:2001vj}).
 The corresponding defect
operators ${\mathcal D}:\HH_{\Gamma_1}\longrightarrow\HH_{\Gamma_2}$ satisfy gluing conditions
\beq\label{defglue}
\left(\begin{array}{c} a_n^2\\\ol{a}_{-n}^2\end{array}\right){\mathcal D}
={\mathcal D}{\mathcal O}\left(\begin{array}{c} a_n^1\\\ol{a}_{-n}^1\end{array}\right)
\eeq
for ${\mathcal O}\in{\rm O}(1,1)$. Here 
the $a_n^i$ and $\ol{a}_n^i$ denote the modes of holomorphic and antiholomorphic $\widehat{\mathfrak u}(1)$-currents in theory $i$. If
the gluing condition ${\mathcal O}$ is diagonal, then it glues together holomorphic and antiholomorphic currents separately, and the resulting
defect is topological.

A gluing condition ${\mathcal O}$ is {\bf admissible} if the lattice
\beq
\Gamma_{\mathcal O}:=\{\gamma\in\Gamma_1\,\big|\,{\mathcal O}\gamma\in\Gamma_2\}\subset\Gamma_1
\eeq
of intertwiners $P_\gamma^{\mathcal O}:\VV_\gamma\subset \HH_1\longrightarrow\VV_{{\mathcal O}\gamma}\subset\HH_2$ implementing the gluing conditions \eq{defglue} has maximal rank.
In this case there are enough intertwiners to construct defect operators
\beq
{\mathcal D}_{\mathcal O}(\chi):=g_{\mathcal O}\sum_{\gamma\in\Gamma_{\mathcal O}}
P_\gamma^{\mathcal O} e^{i\chi(\gamma)}\,,
\eeq
where $\chi\in\Gamma\left(\Gamma_{\mathcal O}\otimes\RR\right)^*$ and
\beq
g_{\mathcal O}=\sqrt{\|\pi_{\mathcal O}(\Gamma_{\mathcal O})\|}
\eeq
is the square root of the volume of the lattice
\beq
\pi_{\mathcal O}(\Gamma_{\mathcal O})=\left\{\left(\begin{array}{c}\pi(\gamma)\\\ol{\pi}({\mathcal O}\gamma)\end{array}\right)\,\Big|\,\gamma\in\Gamma_{\mathcal O}\right\}\,.
\eeq
Here $\pi$ and $\ol{\pi}$ denote the projections on holomorphic and antiholomorphic charges respectively.

In order to describe the intertwiners in more detail, we make use of the folding trick, which 
relates defect operators ${\mathcal D}:\HH_1\longrightarrow\HH_2$ and boundary states
$\|B\keti\in\HH_1^*\otimes\HH_2$ in the folded theory (see \eg \cite{WOng:1994pa,Oshikawa:1996ww,Bachas:2001vj}).
The latter satisfy gluing conditions
\beq\label{bdglue}
\left(
\left(\begin{array}{c} \widehat{a}^1_n \\ a^2_n\end{array}\right)
-S \left(\begin{array}{c} \widehat{\ol{a}}^1_{-n} \\ \ol{a}^2_{-n}\end{array}\right)
\right)\|B\keti=0\,,
\eeq
where $S\in{\rm O}(2)$ and $\widehat{a}_n^1$ and $\widehat{\ol{a}}_n^1$ are the
modes of currents of the folded theory which in terms of the ones in the original theory read\footnote{For convenience of later notation, the currents built out of the $\widehat{a}_n^1$ and $\widehat{\ol{a}}_{n}^1$ are minus the folded currents $j$ and $\ol{\jmath}$ respectively.}
\beq\label{unfold}
\widehat{a}_n^1=\ol{a}_{-n}^1\,,\quad
\widehat{\ol{a}}_n^1=a_{-n}^1\,.
\eeq
Therefore, the gluing conditions in the unfolded and folded theory are related by
\beq
S={1\over{\mathcal O}_{22}}\left(\begin{array}{cc}
-{\mathcal O}_{21} & 1 \\ {\rm det}({\mathcal O}) & {\mathcal O}_{12}\end{array}\right)\,.
\eeq
Ishibashi states implementing the gluing condition \eq{bdglue} can  be written as
\beq
e^{\sum_{n>0}{1\over n}\left(
S_{11}\widehat{a}^1_{-n}\widehat{\ol{a}}^1_{-n}
+S_{12}\widehat{a}_{-n}^1\ol{a}_{-n}^2
+S_{21}a_{-n}^2\widehat{\ol{a}}_{-n}^1
+S_{22}a_{-n}^2\ol{a}^2_{-n}\right)}|(Q_1,Q_2,\ol{Q}_1,\ol{Q}_2)\ket\,,
\eeq
for charges satsifying
\beq
\left(\begin{array}{c} Q_1 \\ Q_2\end{array}\right)=S
\left(\begin{array}{c} \ol{Q}_1 \\ \ol{Q}_2\end{array}\right)\,.
\eeq
Unfolding yields the intertwiners
\beq\label{intertwiners}
P_x^{\mathcal O}=
e^{\sum_{n>0}{1\over n}\left(
S_{11}a^1_{n}\ol{a}^1_{n}
+S_{12}a_{n}^1\ol{a}_{-n}^2
+S_{21}a_{-n}^2\ol{a}_{n}^1
+S_{22}a_{-n}^2\ol{a}^2_{-n}\right)}\,,
\eeq
where monomials in operators $A_i$ of theory $i$ have to be regarded as
\beq
A_1A_2=A_2A_1=A_2|{\mathcal O}x\ket\bra x|A_1\,.
\eeq
For $\mathcal{O}=\mathcal{O}(\theta)$ (\cf \eq{Ofam})
\beq
S=\left(\begin{array}{cc}-\tanh(\theta) & {1\over\cosh(\theta)} \\ {1\over\cosh(\theta)} & \tanh(\theta)\end{array}\right)\,.
\eeq
Furthermore, 
\beq
\pi_{\mathcal O}(\Gamma_1)=\left\{{1\over\sqrt{2}}\left(\begin{array}{c} {n\over R_1}+mR_1 \\
{n\over R_1}e^\theta-mR_1e^{-\theta}\end{array}\right)\,\Big|\,n,m\in\ZZ
\right\}\,,
\eeq
and 
\beq
\|\pi_{\mathcal O}(\Gamma_1)\|^2=\Big|{\rm det}\left(
\begin{array}{cc} {1\over 2R_1^2}(1+e^{2\theta}) & 0 \\ 0 & {R_1^2\over 2}(1+e^{-2\theta})\end{array}
\right)\Big|=\cosh^2(\theta)\,.
\eeq
Hence
\beq
g_{\mathcal O}=\sqrt{\cosh(\theta){\rm ind}_{\Gamma_{\mathcal O}\subset\Gamma_1}}\,.
\eeq
The boundary states \eq{bdglue} are of course just boundary states on a two torus $T^2$ with torus cycles of radius $R_1$ and $R_2$ respectively. So they correspond either to 
D1- or D0/D2-branes on this torus. D1-branes are geometrically characterized by winding numbers
$(k_1, k_2)$ around the two torus cycles; by T-duality they are mapped to bound states of D0- and D2 branes. 

The relation between the data used in this paper and the winding numbers is obtained by identifying
\beq
e^{-\theta}=\tan \vartheta
\eeq
where $\vartheta$ is the angle between the D1-brane and the torus cycle of 
radius $R_1$. In terms of $\vartheta$, the gluing condition $S$ of the boundary condition reads\footnote{Note that the signs for the off-diagonal matrix entries can be changed by twisting
the folding convention \eq{unfold} by $\ZZ_2$ automorphisms of the $\widehat{\mathfrak u}(1)$ current algebra.}
\beq
S = \left( \begin{array}{cc} -\cos 2\vartheta & \sin 2\vartheta \\ \sin 2\vartheta & \cos 2\vartheta \end{array} \right)\,,
\eeq
and for the $g$-factor one obtains
\beq
g_{\cal O} = \sqrt{\frac{k_1k_2}{\sin 2 \vartheta}} = 
\sqrt{\frac{k_1^2 R_1^2+k_2^2R_2^2}{2R_1 R_2}}\,.
\eeq
In the last step we have used that
\beq
\tan \vartheta = \frac{k_2 R_2}{k_1 R_1}  \ .
\eeq
As it should, the $g$-factor is proportional to the length of the D1-brane.
\subsection*{Deformation defects}
In this Section we identify the defects associated to the exactly marginal bulk perturbation
along the family of circle theories. Since this perturbation preserves the holomorphic and anitholomorphic current algebras, also the associated defects should preserve it. Thus, they are among
the class of defects discussed in the previous Section. 
Indeed, as will be argued below, the deformation
from a theory with circle radius $R_2$ to one with $R_1$ is given by the 
$\widehat{\mathfrak u}(1)\oplus \ol{\widehat{\mathfrak u}}(1)$-preserving defect ${\mathcal D}_{\mathcal O}={\mathcal D}_{\mathcal O}(\chi=0)$ with
\beq
{\mathcal O}={\mathcal O}(\theta)\,,\quad e^\theta={R_1\over R_2}\,.
\eeq
In particular, 
\beq
\Gamma_{\mathcal O}=\Gamma_1\,,\quad g_{\mathcal O}=\sqrt{\cosh(\theta)}\,,\quad
S=\left(\begin{array}{cc} -\tanh(\theta)&{1\over\cosh(\theta)}\\{1\over\cosh(\theta)}&\tanh(\theta)\end{array}\right)\,.
\eeq
In terms of boundary conditions in the folded theory, they correspond to D1-branes stretching diagonally across the two-torus
with cycles of radii $R_1$ and $R_2$.

To see that these are the deformation defects, one has to show that perturbing the theory with radius $R_2$ on the half cylinder going to $-\infty$ gives rise to cylinder amplitude with defect ${\mathcal D}_{\mathcal O}$ separating the theory with radius $R_1$ on the half cylinder extending to $-\infty$ from the original theory with radius $R_2$ on the half cylinder extending to $+\infty$. Since we have to compare two families, it is
sufficient to do the comparison to first order in all points in the family, \ie one has to show that $\pi$ times 
the derivative of amplitudes involving the defect ${\mathcal D}_{\mathcal O}$ with respect to 
$\ln(R_1)$ is equal to the first order deformation with $\varphi=j\ol{\jmath}$ on the $R_1$-side of the defect
\beq\label{compeq}
\pi\partial_{\ln(R_1)}\bra\cdots{\mathcal D}_{\mathcal O}\cdots\ket=
\int_{C}{{\rm d}^2z\over -2i}\bra\cdots{\mathcal D}_{\mathcal O}j(z)\ol{\jmath}(\ol{z})\cdots\ket\,.
\eeq
Since we know that the deformation defect preserves the current algebra, 
it is sufficient to show this for the amplitudes defining the action of the defect
operator on highest weight states and the ones encoding the gluing conditions.

For the former consider the cylinder amplitude
\beq
{\mathcal A}_{R_1,R_2}^{\gamma_1,\gamma_2}={}_{R_2}\bra\gamma_2|{\mathcal D}_{\mathcal O}|\gamma_1\ket_{R_1}=g_{\mathcal O}\delta_{\gamma_2,{\mathcal O}\gamma_1}\,.
\eeq
Deriving with respect to $\ln(R_1)$ one obtains\footnote{For $\gamma_1$ we take a constant section in the family, \ie the argument of the $\delta$-function is constant.}
\beq\label{derop}
\partial_{\ln(R_1)}{\mathcal A}_{R_1,R_2}^{\gamma_1,\gamma_2}=\partial_{\theta}{\mathcal A}_{R_1,R_2}^{\gamma_1,\gamma_2}={1\over 2}\tanh(\theta){\mathcal A}_{R_1,R_2}^{\gamma_1,\gamma_2}\,.
\eeq
On the other hand, the first order perturbation calculated on the complex plane with the defect placed on the unit circle is given by
\beq
\partial_\lambda\big|_{\lambda=0}{\mathcal A}_{R_1,R_2}^{\gamma_1,\gamma_2}(\lambda)=
\int_{D_1}{{\rm d}^2z\over -2i}{}_{R_2}\bra\gamma_2|{\mathcal D}_{\mathcal O}j(z)\ol{\jmath}(\ol{z})|\gamma_1\ket_{R_1}\,.
\eeq
Now, using the form \eq{intertwiners} of the intertwiners one easily obtains
\beqn
{}_{R_2}\bra\gamma_2|{\mathcal D}_{\mathcal O}j(z)\ol{\jmath}(\ol{z})|\gamma_1\ket_{R_1}&=&
{}_{R_1}\bra{\mathcal O}^{-1}\gamma_2|e^{\sum_{n>0}{S_{11}\over n}a_n^1\ol{a}_n^1}
\sum_{r,s}a_rz^{-r-1}\ol{a}_s\ol{z}^{-s-1}|\gamma_1\ket_{R_1}\nonumber\\
&=&
\left({Q\ol{Q}\over|z|^2}+{S_{11}\over(1-|z|^2)^2}\right){\mathcal A}_{R_1,R_2}^{\gamma_1,\gamma_2}\,.
\eeqn
Introducing the spacial cutoff $\widetilde{\epsilon}$, the first summand gives a contribution
\beq
I_1=\int_{D_1-D_{\widetilde{\epsilon}}}{{\rm d}^2z\over -2i|z|^2}=-2\pi\ln(\widetilde{\epsilon})\,,
\eeq
whereas the second summand yields
\beq
I_2=\int_{D_{1-\widetilde{\epsilon}}}{{\rm d}^2z\over -2i}{1\over(1-|z|^2)^2}=\pi{(1-\widetilde{\epsilon})^2\over
1-(1-\widetilde{\epsilon})^2}\,.
\eeq
Transforming back to the cylinder $z\mapsto w=\ln(z)$, the cutoff $\widetilde{\epsilon}$ 
on the disk can be expressed in the cutoff $\epsilon$ on the cylinder 
\beq
1-\widetilde{\epsilon}=e^{-2\pi\epsilon}\,.
\eeq
Thus, the contributions to the cylinder are
\beq\label{cylcontr}
I_1=-2\pi\ln(1-e^{-2\pi\epsilon})=O(\epsilon)\,,\qquad
I_2=\pi{e^{-4\pi\epsilon}\over 1-e^{-4\pi\epsilon}}={1\over 4\epsilon}-{\pi\over 2}+O(\epsilon)\,.
\eeq
We use the minimal subtraction scheme, in which renormalization of fields and coupling constants
are defined in such a way that they exactly cancel the singularities $\epsilon^{-n}$, $n>0$ and $\ln(\epsilon)$. Thus, we obtain
\beq
\partial_\lambda\big|_{\lambda=0}{\mathcal A}_{R_1,R_2}^{\gamma_1,\gamma_2}(\lambda)=-{\pi\over 2}S_{11}{\mathcal A}_{R_1,R_2}^{\gamma_1,\gamma_2}
={\pi\over 2}\tanh(\theta){\mathcal A}_{R_1,R_2}^{\gamma_1,\gamma_2}\,.
\eeq
Comparing with \eq{derop} we indeed find that equation \eq{compeq} is satisfied for the amplitudes 
${\mathcal A}_{R_1,R_2}^{\gamma_1,\gamma_2}$ defining the action of the defect operator on highest weight states.

As a next step we  show that the same is true for amplitudes encoding the gluing conditions.
The latter can be probed by cylinder amplitudes of descendents of the vacuum. We will present the calculation for
\beq
G_{R_1,R_2}={}_{R_2}\bra 0|a_1^2{\mathcal D}_{\mathcal O}a_{-1}^1|0\ket_{R_1}
={{\rm det}({\mathcal O})\over {\mathcal O}_{22}}
{}_{R_2}\bra 0|{\mathcal D}_{\mathcal O}|0\ket_{R_1}={1\over\sqrt{\cosh(\theta)}}\,,
\eeq
The calculation for the remaining gluing amplitudes can be performed in a similar fashion.
The derivative of this amplitude with respect to $\ln(R_1)$ is given by
\beq\label{derglue}
\partial_{\ln(R_1)}G_{R_1,R_2}=\partial_{\theta}G_{R_1,R_2}=-{1\over 2}{\sinh(\theta)\over\cosh^2(\theta)}g_{\mathcal O}\,.
\eeq
Again, first order perturbation theory on the disk yields
\beq
\partial_{\lambda}\big|_{\lambda=0}G_{R_1,R_2}=
\int_{D_1}{{\rm d}^2z\over -2i}{}_{R_2}\bra 0|a_1^2{\mathcal D}_{\mathcal O}j(z)\ol{\jmath}(\ol{z})a_{-1}^1|0\ket_{R_1}\,.
\eeq
The correlation function can be easily calculated to be
\beq
{}_{R_2}\bra 0|a_1^2{\mathcal D}_{\mathcal O}j(z)\ol{\jmath}(\ol{z})a_{-1}^1|0\ket_{R_1}=
-{\sinh(\theta)\over\cosh^2(\theta)}g_{\mathcal O}\left(1+{1\over(1-|z|^2)^2}\right).
\eeq
The integral $I_2$ over the second summand has already been calculated (\cf \eq{cylcontr}), and the 
integral over the first summand $1$ of course yields $\pi$. Thus, one obtains
\beq
\partial_{\lambda}\big|_{\lambda=0}G_{R_1,R_2}=-{\pi\over 2}{\sinh(\theta)\over\cosh^2(\theta)}\,,
\eeq
which agrees with $\pi$ times the derivative \eq{derglue}. Hence, the defects ${\mathcal D}_{\mathcal O}$ are indeed the deformation defects along the family of the free bosonic CFTs. 
\subsection*{Casimir energy from boundary conditions}
Next, we calculate explicitly the Casimir energy between deformation defects ${\mathcal D}_{\mathcal O}$
and ${\mathfrak u}(1)$-preserving boundary conditions. The corresponding boundary states
obey gluing conditions
\beq
(a_n-T\ol{a}_{-n})\|B\keti^T=0\,,\quad T\in{\rm O}(1)\cong\ZZ_2\,,
\eeq 
and are given by
\beq
\|B(\chi)\keti^T=g^T\sum_{\gamma\in\Gamma^T}e^{i\chi(\gamma)}
e^{\sum_{n>0}{T\over n}a_{-n}\ol{a}_{-n}}|\gamma\ket\,.
\eeq
The gluing condition $T$ is $+1$ for Dirichlet and $-1$ for Neumann boundary conditions respectively. Moreover,
\beq
\Gamma^T=\Gamma\cap\{(Tx,x)\,|\,x\in\RR\}
\eeq
denotes the lattice of Ishibashi states, 
\beq
g^T=\sqrt{\|\pi(\Gamma^T)\|}
\eeq
 is the square root of the volume of the projection of $\Gamma^T$ on the holomorphic charges, and 
 $\chi\in\left(\Gamma^T\otimes\RR\right)^*$ determines position or Wilson line in case of
  Dirichlet and Neumann boundary conditions, respectively.
  For a free boson on a circle of radius $R$, one obtains 
 \beq
 \Gamma^T={1\over\sqrt{2}R^T}\ZZ\,,\quad g^T={1\over\sqrt{\sqrt{2}R^T}}\,.
 \eeq

The Casimir energy between a defect ${\mathcal D}$ and a boundary condition $\|B\keti$ can be obtained from the amplitude on a semi-infinite cylinder of circumference $U$ with the boundary condition imposed on the finite end of the cylinder, the defect placed a distance $\epsilon$ away from it and the vacuum inserted
at the infinite end of the cylinder (\cf figure \ref{fig2}a)
\beq
{\mathcal B}_{B,{\mathcal D}}^{\epsilon,U}=\brai B\|e^{-2\pi{\epsilon\over U}(L_0+\ol{L}_0)}{\mathcal D}|\Omega\ket\,.
\eeq
Namely,
\beq\label{bosonedef}
{\mathcal E}_{B,{\mathcal D}}^\epsilon=-\lim_{U\to\infty}{1\over U}\ln({\mathcal B}_{B,{\mathcal D}}^{\epsilon,U})\,.
\eeq
For ${\mathfrak u}(1)$-preserving defects and boundary conditions in the free boson theory, this can be easily calculated (\cf \cite{Bachas:2001vj,Bachas:2007td}).
The cylinder amplitude is given by
\beqn\label{bexplicit}
{\mathcal B}_{B^T,{\mathcal D}_{\mathcal O}}^{\epsilon,U}&=&
g_{\mathcal O}g^T{}_{R_2}\bra 0|
e^{\sum_{n>0}{T\over n}a^2_n\ol{a}^2_n}e^{-2\pi{\epsilon\over U}(L_0+\ol{L}_0)}e^{\sum_{n>0}{S_{22}\over n}a_{-n}^2\ol{a}_{-n}^2}|0\ket_{R_2}\\
&=&g_{\mathcal O}g^T\prod_{n>0}{}_{R_2}\bra 0 | 
e^{{T\over n}a^2_n\ol{a}^2_n}e^{-2\pi{\epsilon\over U}(L_0+\ol{L}_0)}e^{{S_{22}\over n}a_{-n}^2\ol{a}_{-n}^2}|0\ket_{R_2}\nonumber\\
&=&g_{\mathcal O}g^T\prod_{n>0}\sum_{l\geq 0}{1\over (l!)^2}\left({TS_{22}e^{-4\pi n{\epsilon\over U}}\over n^2}\right)^l{}_{R_2}\bra 0 |\left(a_n^2\ol{a}_n^2\right)^l\left(a_{-n}^2\ol{a}_{-n}^2\right)^l|0\ket_{R_2}\nonumber\\
&=&g_{\mathcal O}g^T\prod_{n>0}\sum_{l\geq 0}(TS_{22}e^{-4\pi n {\epsilon\over U}})^l\nonumber\\
&=&g_{\mathcal O}g^T\prod_{n>0}{1\over 1-TS_{22}e^{-4\pi n {\epsilon\over U}}}\,.\nonumber
\eeqn
Using the Euler-MacLaurin formula, one finds the exact expression of the Casimir energy 
\beq\label{dilog}
{\mathcal E}_{B^T,{\mathcal D}_{\mathcal O}}^\epsilon={{\rm Li}_2(TS_{22})\over 4\pi\epsilon}\,,
\quad
{\mathcal E}_{B^T,{\mathcal D}_{\mathcal O}}=\epsilon{\mathcal E}_{B^T,{\mathcal D}_{\mathcal O}}^\epsilon
={{\rm Li}_2(TS_{22})\over 4\pi}
\eeq
in terms of the dilogarithm \cite{Bachas:2001vj,Bachas:2007td}. 
\subsection*{Flows induced by boundary conditions}
Since the boundary conditions $B^T$ preserve the $\widehat{\mathfrak u}(1)$ current algebras, they transform along
smoothly under deformations generated by\footnote{This is due to the fact that the perturbing field
$\varphi=j\ol{\jmath}$ via bulk-boundary OPE does not induce relevant perturbations on the boundary.}  $\varphi=j\ol{\jmath}$, and in particular
their gluing conditions $T$ do not change.

Thus, a boundary condition $B^T$ gives rise to a flow on the moduli space of the free boson CFT as described in Section
\ref{sec2}. 
From the exact formula \eq{dilog} it is easy to obtain\footnote{$\partial_x{\rm Li}_2(x)={\ln(1-x)\over x}$}
\beq
\partial_{\ln(R_1)}\big|_{R_1=R_2}{\mathcal E}_{B^T,{\mathcal D}_{\mathcal O}}
=\partial_{\theta}\big|_{\theta=0}{\mathcal E}_{B^T,{\mathcal D}_{\mathcal O}}
=-{T\over 4\pi}\,.
\eeq
This  agrees with the general form \eq{generalgrad} of the first order perturbative calculation of the derivative of $\mathcal{E}$, 
where $B_{j\ol{\jmath}}^{(B^T)}=T$ and $\partial_\lambda=\pi\partial_{\ln(R)}$. 

In fact, one does not need the dilogarithm formula \eq{dilog} to arrive at this result. By means of 
\beqn
\partial_{\ln(R_1)}\big|_{R_1=R_2}{\mathcal E}_{B^T,{\mathcal D}_{\mathcal O}}&=&
\partial_{\theta}\big|_{\theta=0}{\mathcal E}_{B^T,{\mathcal D}_{\mathcal O}}=
-\lim_{U\to\infty}{\epsilon\over U}{\partial_{\theta}{\mathcal B}_{B^T,{\mathcal D}_{\mathcal O}}^{\epsilon,U}\over {\mathcal B}_{B^T,{\mathcal D}_{\mathcal O}}^{\epsilon,U}}\Big|_{\theta=0}\\
&=&-\lim_{U\to\infty}{\epsilon\over  Ug^T}{\partial_{\theta}{\mathcal B}_{B^T,{\mathcal D}_{\mathcal O}}^{\epsilon,U}}\big|_{\theta=0}\,.\nonumber
\eeqn
it can be directly calculated from formula \eq{bexplicit}
\beq
{\partial_{\theta}{\mathcal B}_{B^T,{\mathcal D}_{\mathcal O}}^{\epsilon,U}}\big|_{\theta=0}=
g^Tg_{\mathcal O}T{e^{-4\pi{\epsilon\over U}}\over 1-e^{-4\pi{\epsilon\over U}}}\,,
\eeq
yielding the same result
\beq
\partial_{\ln(R_1)}\big|_{R_1=R_2}{\mathcal E}_{B^T,{\mathcal D}_{\mathcal O}}=
-{T\over 4\pi}\,.
\eeq
The corresponding flow is then given by
\beq
{{\rm d}\over{\rm d} t}\lambda={T\over 4}\,,
\eeq
with solutions
\beq
\lambda(t)={Tt\over 4}\,,\quad R(t)=R(0)e^{\pi Tt\over 4}\,.
\eeq
That means that for Dirichlet boundary conditions ($T=1$) the radius flows to $\infty$, whereas
for Neumann boundary conditions ($T=-1$) the radius flows to $0$. This is expected, after all
we have argued in Section \ref{sec2} that these flows are in general
constant reparametrizations of gradient flows for $\ln(g^T)$, hence decrease $g^T$. Thus,
Dirichlet boundary conditions drive the target space circle radius to $\infty$, whereas Neumann
boundary conditions drive it to $0$. 
\subsection*{Casimir energy from defects}
The Casimir energy between two defects ${\mathcal D}$ and ${\mathcal D}\p$ can be obtained from 
the amplitude 
\beq
{\mathcal B}_{{\mathcal D},{\mathcal D}\p}^{\epsilon, U}=
\bra\Omega|{\mathcal D}e^{-2\pi{\epsilon\over U}(L_0+\ol{L}_0)}{\mathcal D}\p|\Omega\ket
\eeq
on an infinite cylinder of circumference $U$ with the vacuum inserted at both ends and
the two defects placed parallel at distance $\epsilon$ on it (\cf figure \ref{fig2}b).
Namely,
\beq
{\mathcal E}_{{\mathcal D},{\mathcal D}\p}^{\epsilon}=-\lim_{U\to\infty}{1\over  U}\ln(
{\mathcal B}_{{\mathcal D},{\mathcal D}\p}^{\epsilon, U})\,.
\eeq
For two ${\mathfrak u}(1)$-preserving defects ${\mathcal D}={\mathcal D}_{\mathcal O}$ between free boson theories
of radii $R_3$ and $R_2$ and 
${\mathcal D}\p={\mathcal D}_{{\mathcal O}\p}$ between theories with radii $R_2$ and $R_1$
this amplitude can easily be calculated. Indeed, the calculation is analogous to the calculation \eq{bexplicit}
in the boundary case with the result
\beqn
{\mathcal B}_{{\mathcal D},{\mathcal D}\p}^{\epsilon, U}&=&
g_{\mathcal O}g_{{\mathcal O}\p}{}_{R_2}\bra \Omega|
e^{\sum_{n>0}{S_{11}\over n}a^2_n\ol{a}^2_n}e^{-2\pi{\epsilon\over U}(L_0+\ol{L}_0)}e^{\sum_{n>0}{S\p_{22}\over n}a_{-n}^2\ol{a}_{-n}^2}|0\ket_{R_2}\\
&=&g_{\mathcal O}g_{{\mathcal O}\p}\prod_{n>0}{1\over 1-S_{11}S\p_{22}e^{-4\pi n {\epsilon\over U}}}\,.\nonumber
\eeqn
As in the boundary case this leads to the Casimir energy
\beq
{\mathcal E}^\epsilon_{{\mathcal D},{\mathcal D}\p}={{\rm Li}_2(S_{11}S\p_{22})\over
4\pi\epsilon}\,,\quad
{\mathcal E}_{{\mathcal D},{\mathcal D}\p}={{\rm Li}_2(S_{11}S\p_{22})\over
4\pi}\,.
\eeq
\subsection*{Flows defined by defects}
Also $\widehat{\mathfrak u}(1)$-preserving defects ${\mathcal D}_{\mathcal O}$ 
behave smoothly under deformations generated by\footnote{The bulk-defect OPE of the perturbing field $\varphi=j\ol{\jmath}$ does not
produce relevant defect fields.} $\varphi=j\ol{\jmath}$. Thus, a defect ${\mathcal D}_{\mathcal O}$ gives rise to
a flow on the moduli space of the free boson theory on one of its sides.
Taking for ${\mathcal D}\p$ a deformation defect, similarly to the boundary case, 
one easily obtains the derivatives of the Casimir energy
\beq
\partial_{\ln(R_1)}\big|_{R_1=R_2}{\mathcal E}_{{\mathcal D},{\mathcal D}\p}
=\partial_{\theta\p}\big|_{\theta\p=0}{\mathcal E}_{{\mathcal D},{\mathcal D}\p}
=-{S_{11}\over 4\pi}\,.
\eeq
Again this agrees with the general perturbative formula \eq{generalgraddef},
where 
\beq
B_{\varphi}^{({\mathcal D})}=S_{11}=-{{\mathcal O}_{21}\over{\mathcal O}_{22}}\,.
\eeq
In contrast to the boundary case, the gluing condition ${\mathcal O}$ of the defect ${\mathcal D}$ 
is not constant under the deformation of the theory on one side, but is deformed according to\footnote{
This can be seen for instance by fusing the defect with the corresponding deformation defect under which the gluing conditions multiply.}
\beq
{\mathcal O}\mapsto{\mathcal O}{\mathcal O}(\pi\lambda)\,.
\eeq
In particular, the relevant coupling constant $S_{11}$ depends on the deformation parameter.
Writing the undeformed gluing condition as 
\beq
{\mathcal O}=A_{s_1,s_2}{\mathcal O}(\theta)\,,
\eeq
with $s_1,s_2\in\ZZ_2$ and 
\beq
A_{s_1,s_2}=\left(\begin{array}{cc}(-1)^{s_1} & 0 \\ 0 & (-1)^{s_2}\end{array}\right)\,,
\eeq
the deformed gluing condition is given by
\beq
A_{s_1,s_2}{\mathcal O}(\theta(\lambda))=A_{s_1,s_2}{\mathcal O}(\theta+\pi\lambda)\,.
\eeq
Hence,
\beq
S_{11}(\lambda)=-\tanh(\theta(\lambda))=-\tanh(\theta(0)+\pi\lambda)\,.
\eeq
Therefore, the flow defined by the defect ${\mathcal D}_{\mathcal O}$ is given by
\beq
{{\rm d}\over {\rm d}t}\lambda={S_{11}(\lambda)\over 4}=-{1\over 4}\tanh(\theta(0)+\pi\lambda)\,.
\eeq
It can be integrated to
\beq
\sinh(\theta(\lambda))=
\sinh(\theta(0)+\pi\lambda)=\sinh(\theta(0))e^{-\pi{t\over 4}}\,,
\eeq
and one immediately finds that that the gluing condition ${\mathcal O}$ 
flows to the diagonal one in its connected component, which corresponds to a topological defect.
This is expected. After all the flow is a reparametrization of the gradient flow of 
\beq
\ln(g_{\mathcal O})={1\over 2}\ln{\rm ind}_{\Gamma_{\mathcal O}\subset\Gamma_1}+{1\over 2}\ln\cosh(\theta)\,,
\eeq
and $\cosh(\theta)$ of course only has a single critical point, its minimum in $\theta=0$, where $\sinh(\theta=0)=0$.

An `attractor mechanism' for  defects in the compactified free boson has         
already been proposed in \cite{Bachas:2007td}, where the full fusion        
product between two arbitrary symmetry preserving defects was worked out.        
It was noted that the difference in $\ln(g)$ before and after fusion, which was interpreted
as entropy release has the same sign as the Casimir force between the two defects. 
In particular, defects are attracted to each other, when $\ln(g)$ decreases under fusion. 
In this sense, 
the minima of $\ln(g)$, which for the free boson correspond to topological defects were 
suggested as attractors.

Let us also briefly discuss the simultaneous flow on both sides of a defect between one and the same theory. 
For these flows the theories on both sides of the defect remain identical during the flow. The flow equation \eq{gen2flow} in this
case become
\beq
{{\rm d}\over{\rm d}t}\lambda={1\over 4}(S_{11}+S_{22})\,.
\eeq
Here, 
\beq
B_{\varphi_1}^{({\mathcal D})}=S_{11}=-{{\mathcal O}_{21}\over{\mathcal O}_{22}}=-\tanh(\theta)\,,\quad B_{\varphi_2}^{({\mathcal D})}=S_{22}={{\mathcal O}_{12}\over{\mathcal O}_{22}}=(-1)^{s_1+s_2}\tanh(\theta)\,.
\eeq
But under simultaneous deformations, the gluing condition ${\mathcal O}=A_{s_1,s_2}{\mathcal O}(\theta)$ of a defect behaves as
\beq
{\mathcal O}\mapsto {\mathcal O}(-\pi\lambda){\mathcal O}{\mathcal O}(\pi\lambda)
=A_{s_1,s_2}{\mathcal O}(\theta+\pi\lambda(1-(-1)^{s_1+s_2}))\,.
\eeq
Thus, the flow equation becomes
\beq
{{\rm d}\over{\rm d}t}\lambda=-{1\over 4}(1-(-1)^{s_1+s_2})\tanh(\theta+\pi\lambda(1-(-1)^{s_1+s_2}))\,.
\eeq
It can be integrated to
\beq
\sinh(\theta(\lambda))=
\sinh(\theta(0)+\pi\lambda(1-(-1)^{s_1+s_2}))=\sinh(\theta(0))e^{-{\pi t\over 4}(1-(-1)^{s_1+s_2})^2}\,.
\eeq
Hence, there is no flow for even $s_1+s_2$. This is expected, because in this case simultaneous deformation does
not change the gluing condition, and therefore $g$ is constant. For odd $s_1+s_2$ on the other hand, 
$\lambda$ flows to a point, where the gluing condition becomes diagonal, \ie the defect becomes topological.
\section{Supersymmetric version and attractor flows}\label{n=2}
The construction of flows on CFT moduli spaces outlined in Section \ref{sec2} 
of course can be applied to supersymmetric conformal 
field theories as well. Indeed, supersymmetry preserving deformations are more robust
and hence easier to deal with than generic ones. Moreover, $N=(2,2)$ SCFTs
and their moduli spaces are of particular interest  in string theory.

$N=(2,2)$ superconformal field theories admit two classes of supersymmetry preserving
bulk perturbations, those coming from chiral primary fields with the same chirality in the holomorphic and the anti-holomorphic sectors, \ie (chiral,chiral)- and (anti-chiral, anti-chiral)-primary fields, 
and those coming from chiral primaries with opposite chirality, \ie (a,c)- and (c,a)-primary fields. 
For non-linear sigma models these correspond respectively to deformations of the complex structure and the K\"ahler structure of the target space.

While all these perturbations (they will be referred to respectively as (c,c)- and (a,c)-perturbations in the following) preserve the $N=(2,2)$ supersymmetry of the bulk theory, they generically cease to be supersymmetric in the presence of boundaries or defect lines. 

More precisely, supersymmetric boundary conditions and defects come in two classes as well, A-type and B-type, depending on which 
of the $N=2$ superconformal subalgebras of the bulk superconformal algebra they preserve\footnote{Defects can preserve the entire $N=(2,2)$-superconformal algebra, \ie they can be of A-type and of B-type at the same time \cite{Brunner:2007qu}. B-type defects have recently been investigated in the context of Landau-Ginzburg models, where they are easily constructed explicitely \cite{Kapustin:2004df,Brunner:2007qu,Carqueville:2009}}.
It is well known that (c,c)-perturbations preserve supersymmetry on A-type boundary conditions and defects, while they generically destroy supersymmetry on boundary conditions and defects of B-type. 
Analogously, (a,c)-perturbations preserve supersymmetry on B-type boundary condtions and defects,
but generically do not preserve supersymmetry on boundary conditions and defects of A-type. 
(See \cite{Brunner:2009mn,Gaberdiel:2009hk} for details on the boundary case and \cite{Brunner:2007ur} for comments on defects.) This applies in particular to exactly marginal perturbations. 

Given a supersymmetric boundary condition or defect in some $N=(2,2)$ superconformal field theory,
the construction in Section \ref{sec2} gives rise to a direction \eq{generalgrad} in the deformation space of the underlying bulk SCFT. In fact,
for A-type boundary conditions or defects this direction is a (c,c)-direction, whereas for B-type boundary
conditions or defects it is an (a,c)-direction, and thus, the corresponding deformation preserves supersymmetry.
This is due to the fact that the relevant couplings of
(a,c)-/(c,c)-perturbing fields to  A-/B-type boundary conditions or defects vanish. For 
boundary conditions this has already been shown in \cite{Ooguri:1996ck}, and the defect case can be treated similarly. (The argument is sketched in Appendix \ref{contourarg}.)

However, while 
all marginal (c,c)- or (a,c)-perturbations are exactly marginal in the bulk, they can induce relevant perturbations on 
boundaries and defects, even in the case supersymmetry is preserved \cite{Brunner:2009mn,Gaberdiel:2009hk}. This happens at `lines of marginal stability' in moduli space. Away from these lines,  the chosen 
defect or boundary condition behaves smoothly under deformations, and by the construction presented in Section \ref{sec2} gives rise to a flow 
on the (c,c)- or (a,c)-moduli space.

As discussed in Section \ref{sec2}, the 
flows are constant reparametrizations of gradient flows of $\ln(g)$, where the ground state
degeneracy $g$ only depends on the topological charges of the corresponding boundary condition or defect. The latter are well defined for any supersymmetric boundary condition or defect irrespective of conformal invariance. Thus, the gradient flows are well defined on the entire moduli space. 

In the case of non-linear sigma models with Calabi-Yau target manifold, 
A-type boundary conditions corresponding to A-branes on Lagrangian submanifolds give 
rise to flows on the complex structure moduli space of the target space. 
B-type boundary conditions, which are given by coherent sheaves on the target space on the other hand induce flows on the K\"ahler moduli space. 

Viewed in the context of string compactifications, the resulting theory in four dimensions is an ${N}=2$ supergravity theory where (for type IIB) the (a,c)-moduli become scalars in the hypermultiplets, and the (c,c)-moduli in the vector multiplets. D-branes can be viewed as black holes of this supergravity theory provided that the boundary conditions are Dirichlet in the uncompactified directions. The black hole solution a priori depends on the background moduli as well as the electric and magnetic charges they carry. The latter are determined by the topological charges of the A-type boundary condition in the internal theory, which can be 
expressed in terms of the homology class, the corresponding A-brane represents in the compactification space.
The values of the moduli at the horizon of the black hole can be obtained as fixed points of a flow
on the respective moduli space \cite{Ferrara:1995ih}.
This flow is referred to as attractor flow in supergravity.
It can be realized as the gradient flow of the logarithm 
$\ln(|Z|^2)$
of the absolute value squared of the
central charge of the corresponding low energy BPS particle (see \eg \cite{Moore:1998pn,Denef:2000nb}). 

The mass of a BPS particle in four dimensions equals the absolute value $|Z|$ of the central charge. Since the graviton vertex operator is the identity in the compactified dimensions, the mass of a BPS particle is given by the g-function of the corresponding boundary condition in the internal theory. As a consequence the absolute value of the central charge is equal to the g-function\footnote{The phase of the central charge is encoded in the boundary condition of the spectral flow operator, from which one constructs the space-time supercharges.}
\beq
|Z|=g_B\,.
\eeq
Therefore, the attractor flow is a constant reparametrization of the gradient flow of $\ln(g_B)$.
The construction of flows by means of Casimir energies presented in Section \ref{sec2} therefore provides a world sheet realization
of the attractor flow encountered in supergravity. Besides, it also generalizes the attractor flow in various directions. For instance, it associates flows not only to D-branes, but also to defects. Moreover, it does not require supersymmetry, nor any target space interpretation.
\subsection*{Acknowledgements}
D.~R.~ is supported by a DFG research fellowship and partially by DOE-grant DE-FG02-96ER40949. I.~B.~ is supported by a EURYI award. We would like to thanks N.~Carqueville, A.~Collinucci and S.~Nampuri for discussions.
\appendix
\section{Perturbative variation of charges}\label{pertapp} 
It is well known that bilinear combinations of ${\mathfrak u}(1)$-currents generate
exactly marginal perturbations (see \eg \cite{chaud}). 
To determine the first order variations
of ${\mathfrak u}(1)$-charges under such deformations we introduce 
a space-time cutoff $\epsilon$ and use the minimal subtraction scheme with respect to this cutoff.
That means the renormalizations of fields and coupling constants are chosen in such a way
that they cancel exactly the singularities proportional to $\ln(\epsilon)$ and $\epsilon^{-n}$, $n>0$.
(For more details on this see \eg the Appendix of \cite{fr}.)

The variation of the ${\mathfrak u}(1)$-charges can be determined by calculating the first order
perturbation of the correlation function
\beq
\bra(Q,\ol{Q})|j(z)|(Q,\ol{Q})\ket={Q\over z}\bra(Q,\ol{Q})|(Q,\ol{Q})\ket={Q\over z}\,.
\eeq
One obtains
\beqn
\partial_{\lambda}\big|_{\lambda=0} \bra(Q,\ol{Q})|j(z)|(Q,\ol{Q})\ket
&=&{1\over -2i} \int_{\PP^1-D_{\epsilon}(0)-D_{\epsilon}(z)}\!\!\!\!\!\!\!\!\!\!\!\!\!\!\!\!\!\!\!\!\!\!\!\!\!\!\!\!
{\rm d}^2 w 
\bra(Q,\ol{Q})|j(w)\ol{\jmath}(\ol{w})j(z)|(Q,\ol{Q})\ket\\
&=&-{\ol{Q}\over 2i} \int_{\PP^1-D_{\epsilon}(0)-D_{\epsilon}(z)}\!\!\!\!\!\!\!\!\!\!\!\!
{{\rm d}^2 w} 
\left({1\over (z-w)^2\ol{w}}+{Q^2\over |w|^2z}\right)\,,\nonumber
\eeqn
where $D_{\epsilon}(z)$ is the disk with radius $\epsilon$ around $z$. 
The integral over the second summand gives rise to a term proportional to $\ln(\epsilon)$
which is compensated by a field redefinition of $|(Q,\ol{Q})\ket$. The first term can be rewritten by
means of 
\beq
{1\over (z-w)^2\ol{w}}=\partial_w{1\over (z-w)\ol{w}}\,,
\eeq
which holds outside the singularities. Thus, one obtains
\beqn
\partial_{\lambda}\big|_{\lambda=0} \bra(Q,\ol{Q})|j(z)|(Q,\ol{Q})\ket
&=&-{\ol{Q}\over 2i} \int_{\PP^1-D_{\epsilon}(0)-D_{\epsilon}(z)}{\rm d}{{\rm d}\ol{w}\over (z-w)\ol{w}}\\
&=&{\ol{Q}\over 2i} \int_{\mathbb{S}^1_\epsilon(0)+\mathbb{S}^1_\epsilon(z)} {{\rm d}\ol{w}\over (z-w)\ol{w}}\nonumber\\
&=&-{\ol{Q}\over 2i}\int_{\mathbb{S}^1_\epsilon(0)}{{\rm d}w\over w(z-w)}
+{\ol{Q}\over 2i}\int_{\mathbb{S}^1_\epsilon(z)}{{\rm d}\ol{w}\over \ol{w}(z-w)}\nonumber\\
&=&-{\pi{\ol Q}\over z}\,,\nonumber
\eeqn
where in the last step, the second summand is zero in the limit $\epsilon\rightarrow 0$.
Hence, 
$\partial_{\lambda} Q=-\pi{\ol Q}$,
and in the same way, one arrives at 
$\partial_{\lambda} \ol{Q}=-\pi{Q}$. 
\section{The Hessian of $\ln(g_{\mathcal D})$ for topological defects}\label{Hessian}
Since $\ln(g_{\mathcal D})$ is critical in topological defects, in those points we have
\beqn
\partial_{\lambda^i}\partial_{\lambda^j}\Big|_{\lambda=0}\ln(g_{\mathcal D})&=&{1\over g_{\mathcal D}}
\partial_{\lambda^i}\partial_{\lambda^j}\Big|_{\lambda=0}g_{\mathcal D}\\
&=&{1\over g_{\mathcal D}}\int{{\rm d}^2z\over -2i}\int{{\rm d}^2w\over -2i}\bra \varphi_i(w,\ol{w})\varphi_j(z,\ol{z})\ket_{\mathcal D}\,,\nonumber
\eeqn
where the correlation function is one on the complex plane with defect ${\mathcal D}$ placed on the unit circle,
and the integrals are over the unit disk with regularization inherited from the cylinder geometry.

Since the defect is topological, the correlation function is given by
\beq
\bra \varphi_i(w,\ol{w})\varphi_j(z,\ol{z})\ket_{\mathcal D}={g_{\mathcal D}g_{ij}\over |z-w|^4}\,.
\eeq
The integral 
\beq
I:=\int{{\rm d}^2z\over -2i}\int{{\rm d}^2w\over -2i}{1\over |z-w|^4}\,,
\eeq
can be calculated. Expansion in the regularization parameter $\epsilon$ yields the finite 
part $I_{\rm reg}={\pi^2\over 4}$. 
Hence,
\beq
\partial_{\lambda^i}\partial_{\lambda^j}\Big|_{\lambda=0}\ln(g_{\mathcal D})={\pi^2\over 4}g_{ij}\,,
\eeq
which for unitary theories is positive definite.

\section{Vanishing of bulk-boundary/-defect couplings for $N=2$ theories}\label{contourarg}
Here we briefly sketch why the couplings of (a,c)-perturbing fields to A-type boundaries and defects, and analogously
the couplings of (c,c)-perturbing fields to B-type boundaries and defects vanish. The argument for the case of 
boundary conditions has already been given in \cite{Ooguri:1996ck}. It can be easily adapted to the 
treatment of defects.
Let us discuss the bulk-boundary coupling of an (a,c)-perturbing field $\varphi$ to an A-type boundary vanishes.  The other cases can be dealt with analogously. 

The perturbing field $\varphi$ is a descendant of an (a,c)-primary $\chi$, \ie
\beq
\varphi=G_{-{1\over 2}}^+\ol{G}_{-{1\over 2}}^-\chi\,,\quad{\rm with}\quad
G^-_{-{1\over 2}}\chi=0=\ol{G}^+_{-{1\over 2}}\chi\,.
\eeq
Here $G^{\pm}$ and $\ol{G}^{\pm}$ are the holomorphic and anti-holomorphic supercurrents respectively. Now, consider the one-point function 
\beq
\bra\varphi(0)\ket_A=g_AB_{\varphi}^{(A)}
\eeq
of $\varphi$ on the unit disk with an A-type boundary condition $A$ imposed on the unit circle.
The action of $G^+$ on $\chi$ can be written as a contour integral
\beq
\left(G_{-{1\over 2}}^+\ol{G}_{-{1\over 2}}^-\chi\right)(0)=\oint{{\rm d}z\over 2\pi i} G^+(z)\left(\ol{G}_{-{1\over 2}}^-\chi\right)(0)
\eeq
around $0$ inside the unit disk. The contour can be deformed to the boundary, where A-type gluing conditions require $G^+=\ol{G}^-$. Thus,
\beqn
\bra\varphi(0)\ket_A&=&\oint{{\rm d}z\over 2\pi i}\bra G^+(z)\left(\ol{G}_{-{1\over 2}}^-\chi\right)(0)\ket_A
=-\oint{{\rm d}\ol{z}\over 2\pi i}\bra \ol{G}^-(\ol{z})\left(\ol{G}_{-{1\over 2}}^-\chi\right)(0)\ket_A\\
&=&\bra \left(\left(\ol{G}_{-{1\over 2}}^-\right)^2\chi\right)(0)\ket_A\,,\nonumber
\eeqn
which vanishes because
$ \left(\ol{G}_{-{1\over 2}}^-\right)^2=0$. This shows that the coupling
$B_{\varphi}^{(A)}$ of an (a,c)-perturbing field $\varphi$ to an A-type boundary condition is zero.

%
%
%
%
\bibliographystyle{pdef}
\bibliography{pdef}

\providecommand{\href}[2]{#2}\begingroup\raggedright\begin{thebibliography}{10}

\bibitem{Brunner:2007ur}
I.~Brunner and D.~Roggenkamp, ``{Defects and Bulk Perturbations of Boundary
  Landau-Ginzburg Orbifolds},'' {\em JHEP} {\bf 04} (2008) 001,
\href{http://arXiv.org/abs/0712.0188}{{\tt 0712.0188}}.

\bibitem{Brunner:2008fa}
I.~Brunner, H.~Jockers, and D.~Roggenkamp, ``{Defects and D-Brane
  Monodromies},''
\href{http://arXiv.org/abs/0806.4734}{{\tt 0806.4734}}.

\bibitem{Bachas:2001vj}
C.~Bachas, J.~de~Boer, R.~Dijkgraaf, and H.~Ooguri, ``Permeable conformal walls
  and holography,'' {\em JHEP} {\bf 06} (2002) 027,
\href{http://arXiv.org/abs/hep-th/0111210}{{\tt hep-th/0111210}}.

\bibitem{Petkova:2000ip}
V.~B. Petkova and J.~B. Zuber, ``Generalised twisted partition functions,''
  {\em Phys. Lett.} {\bf B504} (2001) 157--164,
\href{http://arXiv.org/abs/hep-th/0011021}{{\tt hep-th/0011021}}.

\bibitem{Frohlich:2006ch}
J.~Fr{\"o}hlich, J.~Fuchs, I.~Runkel, and C.~Schweigert, ``Duality and defects
  in rational conformal field theory,'' {\em Nucl. Phys.} {\bf B763} (2007)
  354--430,
\href{http://arXiv.org/abs/hep-th/0607247}{{\tt hep-th/0607247}}.

\bibitem{Bachas:2007td}
C.~Bachas and I.~Brunner, ``{Fusion of conformal interfaces},'' {\em JHEP} {\bf
  02} (2008) 085,
\href{http://arXiv.org/abs/arXiv:0712.0076 [hep-th]}{{\tt arXiv:0712.0076
  [hep-th]}}.

\bibitem{Ferrara:1995ih}
S.~Ferrara, R.~Kallosh, and A.~Strominger, ``{N=2 extremal black holes},'' {\em
  Phys. Rev.} {\bf D52} (1995) 5412--5416,
\href{http://arXiv.org/abs/hep-th/9508072}{{\tt hep-th/9508072}}.

\bibitem{Ferrara:1996um}
S.~Ferrara and R.~Kallosh, ``{Universality of Supersymmetric Attractors},''
  {\em Phys. Rev.} {\bf D54} (1996) 1525--1534,
\href{http://arXiv.org/abs/hep-th/9603090}{{\tt hep-th/9603090}}.

\bibitem{Ferrara:1996dd}
S.~Ferrara and R.~Kallosh, ``{Supersymmetry and Attractors},'' {\em Phys. Rev.}
  {\bf D54} (1996) 1514--1524,
\href{http://arXiv.org/abs/hep-th/9602136}{{\tt hep-th/9602136}}.

\bibitem{Ferrara:1997tw}
S.~Ferrara, G.~W. Gibbons, and R.~Kallosh, ``{Black holes and critical points
  in moduli space},'' {\em Nucl. Phys.} {\bf B500} (1997) 75--93,
\href{http://arXiv.org/abs/hep-th/9702103}{{\tt hep-th/9702103}}.

\bibitem{Moore:1998pn}
G.~W. Moore, ``{Arithmetic and attractors},''
\href{http://arXiv.org/abs/hep-th/9807087}{{\tt hep-th/9807087}}.

\bibitem{borcea}
C.~Borcea, ``Calabi-yau threefolds and complex multiplication,'' in {\em Essays
  on Mirror Manifolds}, S.-T. Yau, ed.
\newblock International Press, 1992.

\bibitem{dr}
D.~Roggenkamp, ``Defects, rationality and complex multiplication.''
  unpublished.

\bibitem{Gukov:2002nw}
S.~Gukov and C.~Vafa, ``{Rational conformal field theories and complex
  multiplication},'' {\em Commun. Math. Phys.} {\bf 246} (2004) 181--210,
\href{http://arXiv.org/abs/hep-th/0203213}{{\tt hep-th/0203213}}.

\bibitem{Fischler:1986tb}
W.~Fischler and L.~Susskind, ``{Dilaton Tadpoles, String Condensates and Scale
  Invariance. 2},'' {\em Phys. Lett.} {\bf B173} (1986)
262.

\bibitem{Fischler:1986ci}
W.~Fischler and L.~Susskind, ``{Dilaton Tadpoles, String Condensates and Scale
  Invariance},'' {\em Phys. Lett.} {\bf B171} (1986)
383.

\bibitem{Keller:2007nd}
C.~A. Keller, ``{Brane backreactions and the Fischler-Susskind mechanism in
  conformal field theory},'' {\em JHEP} {\bf 12} (2007) 046,
\href{http://arXiv.org/abs/0709.1076}{{\tt 0709.1076}}.

\bibitem{WOng:1994pa}
E.~Wong and I.~Affleck, ``{Tunneling in quantum wires: A Boundary conformal
  field theory approach},'' {\em Nucl. Phys.} {\bf B417} (1994)
403--438.

\bibitem{Oshikawa:1996ww}
M.~Oshikawa and I.~Affleck, ``{Defect Lines in the Ising Model and Boundary
  States on Orbifolds},'' {\em Phys. Rev. Lett.} {\bf 77} (1996) 2604--2607,
\href{http://arXiv.org/abs/hep-th/9606177}{{\tt hep-th/9606177}}.

\bibitem{chaud}
S.~Chaudhuri and J.~A. Schwartz, ``{A criterion for integrably marginal
  operators},'' {\em Phys. Lett.} {\bf B219} (1989)
291.

\bibitem{Brunner:2007qu}
I.~Brunner and D.~Roggenkamp, ``{B}-type defects in {L}andau-{G}inzburg
  models,'' {\em JHEP} {\bf 08} (2007) 093,
\href{http://arXiv.org/abs/arXiv:0707.0922 [hep-th]}{{\tt arXiv:0707.0922
  [hep-th]}}.

\bibitem{Kapustin:2004df}
A.~Kapustin and L.~Rozansky, ``On the relation between open and closed
  topological strings,'' {\em Commun. Math. Phys.} {\bf 252} (2004) 393--414,
\href{http://arXiv.org/abs/hep-th/0405232}{{\tt hep-th/0405232}}.

\bibitem{Carqueville:2009}
N.~Carqueville and I.~Runkel, ``On the monoidal structure of matrix
  bi-factorisations,'' \href{http://arXiv.org/abs/arXiv:0909.4381}{{\tt
  arXiv:0909.4381}}.

\bibitem{Brunner:2009mn}
I.~Brunner, M.~R. Gaberdiel, S.~Hohenegger, and C.~A. Keller, ``{Obstructions
  and lines of marginal stability from the world-sheet},'' {\em JHEP} {\bf 05}
  (2009) 007,
\href{http://arXiv.org/abs/0902.3177}{{\tt 0902.3177}}.

\bibitem{Gaberdiel:2009hk}
M.~R. Gaberdiel and S.~Hohenegger, ``{Manifestly Supersymmetric RG Flows},''
\href{http://arXiv.org/abs/0910.5122}{{\tt 0910.5122}}.

\bibitem{Ooguri:1996ck}
H.~Ooguri, Y.~Oz, and Z.~Yin, ``{D-branes on Calabi-Yau spaces and their
  mirrors},'' {\em Nucl. Phys.} {\bf B477} (1996) 407--430,
\href{http://arXiv.org/abs/hep-th/9606112}{{\tt hep-th/9606112}}.

\bibitem{Denef:2000nb}
F.~Denef, ``{Supergravity flows and D-brane stability},'' {\em JHEP} {\bf 08}
  (2000) 050,
\href{http://arXiv.org/abs/hep-th/0005049}{{\tt hep-th/0005049}}.

\bibitem{fr}
S.~F{\"o}rste and D.~Roggenkamp, ``{Current current deformations of conformal
  field theories, and WZW models},'' {\em JHEP} {\bf 05} (2003) 071,
\href{http://arXiv.org/abs/hep-th/0304234}{{\tt hep-th/0304234}}.

\end{thebibliography}\endgroup
\end{document}